\UseRawInputEncoding 
\documentclass[prb,preprint,showpacs,superscriptaddress,longbibliography]{revtex4-1}

\usepackage{amsmath}
\usepackage{graphicx}
\usepackage{amssymb}
\usepackage{times}
\usepackage{color}
\usepackage{multirow}
\definecolor{lr}{rgb}{1.0,0.3,0.3}
\definecolor{dg}{rgb}{0.0,0.5,0.0}

\usepackage{tabularx} 

\usepackage{comment}

\graphicspath{./}
 


\begin{document}

\title{Low-field microwave-free sensors using dipolar spin relaxation of quartet spin states in silicon carbide }

\author{Oscar Bulancea-Lindvall} 
\affiliation{Department of Physics, Chemistry and Biology, Link\"oping University, SE-581 83 Link\"oping, Sweden}
  
\author{Matthew T. Eiles} 
\affiliation{Max-Planck-Institut f\"{u}r Physik komplexer Systeme, N\"{o}thnitzer Str.\ 38, D-01187 Dresden, Germany}

\author{Nguyen Tien Son} 
\affiliation{Department of Physics, Chemistry and Biology, Link\"oping University, SE-581 83 Link\"oping, Sweden}

\author{Igor A. Abrikosov} 
\affiliation{Department of Physics, Chemistry and Biology, Link\"oping University, SE-581 83 Link\"oping, Sweden}

\author{Viktor Iv\'ady}
\email{viktor.ivady@liu.se}
\affiliation{Department of Physics, Chemistry and Biology, Link\"oping University, SE-581 83 Link\"oping, Sweden}
\affiliation{Max-Planck-Institut f\"{u}r Physik komplexer Systeme, N\"{o}thnitzer Str.\ 38, D-01187 Dresden, Germany}
\affiliation{Wigner Research Centre for Physics,  PO Box 49, H-1525, Budapest, Hungary}

\date{\today}

\begin{abstract}

Paramagnetic defects and nuclear spins are the major sources of magnetic field-dependent spin relaxation in point defect quantum bits.   The detection of related optical signals has led to the development of advanced relaxometry applications with high spatial resolution.  The nearly degenerate quartet ground state of the silicon vacancy qubit in silicon carbide (SiC) is of special interest in this respect, as it gives rise to relaxation rate extrema at vanishing magnetic field values and emits in the first near-infra-red transmission window of biological tissues, providing an opportunity for developing novel sensing applications for medicine and biology.  However,  the relaxation dynamics of the silicon vacancy center in SiC have not yet been fully explored.  In this paper, we present results from a comprehensive theoretical investigation of the dipolar spin relaxation of the quartet spin states in various local spin environments.  We discuss the underlying physics and quantify the magnetic field and spin bath dependent relaxation time $T_1$.  Using these findings we demonstrate that the silicon vacancy qubit in SiC can implement microwave-free low magnetic field quantum sensors of great potential.

\end{abstract}
\maketitle


\section{Introduction}

Due to their robustness,  sensitivity and versatility,  point defect quantum bits exhibit enormous potential for quantum sensing.  The stringent requirements of emerging multidisciplinary applications,  for example room temperature operation, sensing at low or zero magnetic field, microwave-free and all-optical control,  and bioinertness, pose numerous challenges for such devices. Novel point defect sensors which can meet some or all of these criteria are continuously sought after. 

The NV center in diamond\cite{DohertyNVreview} is the leading contender in quantum sensing applications\cite{barry_sensitivity_2020} realized by optically addressable point defect qubits. \cite{degen_quantum_2017}  Recent developments in NV relaxometry\cite{tetienne_spin_2013,schmid-lorch_relaxometry_2015, van_der_sar_nanometre-scale_2015,hall_detection_2016,rendler_optical_2017,finco_imaging_2021},  where the variation of the longitudinal spin relaxation time is detected by optical means,  have made high temperature microwave-free sensing applications possible.  Sensing at low magnetic field with such sensors is, however,  not possible due to the large zero-field-splitting of the triplet ground state,  which requires a bias field of approximately 100~mT to bring the spin states to near degeneracy.  The presence of the bias field is often undesirable as it may perturb the sample and influence the measurement.\cite{zheng_zero-field_2019,lenz_magnetic_2020,fu_sensitive_2020,wang_zero-field_2021}

Relaxometry-based sensing has not been explored for other point defect qubits, such as the divacancy\cite{koehl_room_2011,bulancea-lindvall_dipolar_2021} and the negatively charged silicon vacancy\cite{SoltamovPhysRevLett2012,widmann_coherent_2015,Simin2016} in silicon carbide (SiC). \cite{bulancea-lindvall_dipolar_2021} The latter defect is, however,  attractive for low magnetic field relaxometry applications owing to its small zero-field splitting value and  resulting quasi degenerate electron spin states.  For such applications, a detailed understanding of the relaxation processes of this defect is crucial.

The negatively charged silicon-vacancy in SiC provides an optically addressable  point defect whose quartet ground state spin\cite{Riedel2012} has a long room temperature coherence time\cite{Widmann2014,simin_locking_2017}.  The unusual high spin state has been utilized in various applications including quantum sensing\cite{Lee2015, Simin2016, Niethammer2016,Anisimov2016},  room temperature maser\cite{Kraus2014},  and near infra-red quantum information processing\cite{nagy_high-fidelity_2019,son_developing_2020,babin_fabrication_2022}.  In recent years,  considerable attention has been paid to experimental characterization of  the relaxation dynamics of the quartet silicon vacancy spin states in a broad temperature range in various SiC samples.\cite{simin_locking_2017,soltamov_excitation_2019,ramsay_relaxation_2020} The external field dependence of the relaxation processes have received little attention thus far,  but must be understood in detail for various relaxometry applications.  Recent theoretical developments have enabled parameter-free calculations of the two major contributions to the longitudinal spin relaxation, namely the temperature dependent spin-lattice relaxation\cite{gugler_ab_2018,park_spin-phonon_2020,xu_spin-phonon_2020} and the magnetic field dependent dipolar spin relaxation induced by local environmental spins\cite{IvadyPRb2020,ivady_photoluminescence_2021,bulancea-lindvall_dipolar_2021}.

In this paper, we study the magnetic field and local spin environment dependence of the dipolar spin relaxation of the V1 and V2  silicon vacancy centers in the 4H polytype of SiC (4H-SiC). \cite{IvadyVSi-4H} We consider various environmental spin species,  such as the naturally abundant $^{13}$C and $^{29}$Si nuclear spins and  spin-1/2 and spin-1 point defects over a wide range of concentrations.  We  identify the most relevant level anti-crossings (LACs), where environmental spins efficiently relax the quartet spin states, and quantify the spin relaxation time $T_1$.  We find several narrow resonances, where the spin relaxation time can vary over several orders of magnitude within a small magnetic field interval close to $B = 0$.  Utilizing these observations we propose novel quantum sensors for biological use and estimate their sensitivity.

The paper is organized as follows.  In section~\ref{sec:meth}, we detail the models and the simulation technique used in this article.   In section~\ref{sec:res}, we present our computational results, while in section~\ref{sec:disc} we discuss our proposal for low-field refractometry with the silicon vacancy center in SiC.  Finally,  in section~\ref{sec:sum}, we summarize our findings.

\section{Methodology}
\label{sec:meth}

We model the relaxation dynamics of many-spin systems consisting of a quartet silicon vacancy electron spin and a number environmental spins of different kinds.  We consider spin-1/2 $^{13}$C and $^{29}$Si nuclear spins and doublet and triplet electron spin environments.  More information on relevant paramagnetic defects in SiC can be found in Ref.~[\onlinecite{bulancea-lindvall_dipolar_2021}].   Nuclear and electron baths are considered independently and calculated separately.  

The spin Hamiltonians of the many-body system is given by
\begin{equation}\label{eq:1}
H  =  H_{0} + \sum_{i} H^{\text{x}}_{i}  + \sum_i  H_{0i}^{\text{x}} + \sum_{i,j}  H_{ij}  \text{,}
\end{equation} 
where the Hamiltonian $H_0$ of the quartet silicon vacancy spin includes the zero-field splitting (ZFS) and the Zeeman term (ZE),
\begin{equation} \label{eq:2}
H_0 = H_{\text{ZFS}}  + H_{\text{ZE}} =  D \left( S_z^2 - \frac{5}{4} \right) + \left( g_{e} S_z + g_{3\parallel} \frac{S_{+}^3 - S_{-}^3 }{4i} \right) \mu_B B_z  \text{.}
\end{equation}
In Eq.~(\ref{eq:2}), the ZFS parameter $D = 2.6$~MHz ($D = 35.0$~MHz) for the V1 center (V2 center) in 4H-SiC\citep{IvadyVSi-4H}.  The second term on the r.h.s.\ of Eq.~(\ref{eq:2}) accounts for the linear and nonlinear Zeeman interactions of the silicon vacancy center, where $g_{e} = 2.0$, $g_{3\parallel} = 0.6$, and $\mu_B$ is the Bohr magneton.\cite{Simin2016} The non-linear Zeeman term is a direct consequence of the three-fold rotation symmetry and the high spin state of the defect.\citep{IvadyNPJCM2018,Simin2016}

The Hamiltonian $H^{\text{x}}_i$ in Eq.~(\ref{eq:1}) depends on the considered spin bath and accordingly $\text{x} = \left\lbrace \text{n}, \text{de}, \text{te}\right\rbrace$.  For nuclear spins, it includes only the nuclear Zeeman interaction term,
\begin{equation}
H_{i}^{\text{n}} = -g_{\text{N}} \mu_{N} I_{i,z} B_z \text{,}
\end{equation}
 where $g_{\text{N}}$ is the nuclear g-factor of either the $^{13}$C or the  $^{29}$Si isotope,  $\mu_{N}$ is the nuclear magneton,  and $I_{i,z}$ is the doublet nuclear spin $z$ operator.  For doublet electron (de) spins,  the Hamiltonian includes only the linear Zeeman term,
\begin{equation}
H_{i}^{\text{de}} = g_{e} \mu_{B} S_{i,z} B_z \text{,}
\end{equation} 
while for triplet electron (te) spins the Hamiltonian includes the ZFS and the linear Zeeman term, 
\begin{equation}
H_{i}^{\text{te}} =  D_i \left( S_{i,z}^2 - \frac{1}{3} \right) + \frac{E_i}{2} \left( S_{+}^2 + S_{-}^2 \right) + g_{e} \mu_{B} S_{i,z} B_z \text{,}
\end{equation} 
where $D_i$ and $E_i$ are ZFS parameters of triplet spin defect $i$.  In our study,  we the triplet spin bath consists divacancy defects, whose ZFS parameters can be found in Ref.~[\onlinecite{falk_polytype_2013}]. 

The Hamiltonian $H_{0i}^{\text{x}}$ terms in Eq.~(\ref{eq:1}) account for the interaction between the silicon vacancy center and the environmental spins.  For nuclear spins the coupling term can be written as,
\begin{equation}
H_{0i}^{\text{n}} = \mathbf{S}A_i \mathbf{I}_i \text{,}
\end{equation}
where $ \mathbf{S}$ and $\mathbf{I}_i$ are the quartet electron spin and spin-1/2 nuclear spin vector operators, respectively,  and $A_i$ is the hyperfine tensor determined from ab initio density functional theory (DFT) calculations.  The details of the DFT hyperfine calculations can be found in Ref.~[\onlinecite{IvadyPRb2020}].  For an electron spin bath, the coupling Hamiltonian term is equal to the magnetic dipole-dipole coupling and can be written as
\begin{equation}
H_{0i}^{\text{de}} = H_{0i}^{\text{te}} = - \frac{\mu_0 g_e^2 \mu_B^2}{4 \pi r^3} \left( 	3 \left( \mathbf{S} \mathbf{r} \right) \left( \mathbf{S}_i \mathbf{r} \right)  -  \left( \mathbf{S} \mathbf{S}_i \right)\right) \text{,}
\end{equation}
where $\mu_0$ is the vacuum permeability,  $\mathbf{S}_i$ is either the double or the triplet electron spin operator vector of environmental spin defect $i$,  $\mathbf{r}$ is the distance vector of the silicon vacancy and the paramagnetic defect, and $r = \left| \mathbf{r} \right|$.  Finally, the last term on the right hand side of Eq.~(\ref{eq:1}) accounts for intra-spin bath couplings.  According to our numerical tests, this term does not significantly contribute to spin relaxation phenomena and it is neglected hereinafter. 

In the numerical simulations we investigate the relaxation dynamics of a central electron spin by utilizing a clustering-based computational method recently developed in Ref.~[\onlinecite{IvadyPRb2020}] and summarized in Ref.~[\onlinecite{bulancea-lindvall_dipolar_2021}]. This method uses an extended Lindbladian to facilitate effective interactions.   In this work,  we use first-order cluster approximation,  i.e.\  a system of $N$ spins is divided into $N$ subsystems.  Each subsystem includes the quartet silicon vacancy electron spin and one spin from the surrounding spin bath. The spin Hamiltonian of the first order cluster systems can be written as
\begin{equation} \label{eq:hamCA1}
\widetilde{H}_i^{1} = H_{0} +  H_{i}  +  H_{0i}   \text{.}
\end{equation} 

Our spin bath models include either $N = 31$ (spin-1/2 or spin-1) paramagnetic defects or $N = 127$ ($^{13}$C or $^{29}$Si) nuclear spins.  According to our previous calculations in 4H-SiC\cite{bulancea-lindvall_dipolar_2021} and diamond\cite{IvadyPRb2020,ivady_photoluminescence_2021},  these values ensures that our simulations are converged with respect to the bath size.   In order to obtain ensemble averaged quantities we carry out averaging over different  spin bath configurations.  In all cases,  an ensemble of 200 random spin bath configurations are considered.  All configurations  correspond to either a given isotope abundance or a given electron spin defect concentration.  We note that in the nuclear spin bath calculations we do not consider those random configurations that contain $^{13}$C nuclear spin in the first neighbour shell of the defect.  In the electron spin bath calculations we exclude those spins whose coupling strength exceed 100~MHz,  since in such cases the energy level structure is completely mixed by the strong  interaction and the center does not function as a regular silicon vacancy qubit. These restrictions affect only a marginal part of the random ensemble.   

For the different spin environments, we carry out  two different time-dependent studies.  First, starting from a highly polarized state of the quartet spin and a thermal state of the bath, we simulate the time propagation of the many-spin system over a short period of time,  $t = 1$~$\mu$s,  and study the amount of population transferred from the initial state to the other states of the quartet silicon vacancy.  These qualitative studies shed light on the external parameter dependence of the  silicon vacancy-environment couplings. Furthermore,  the results of such calculations may be compared with photoluminescence (PL) studies.\cite{bulancea-lindvall_dipolar_2021,ivady_photoluminescence_2021}  Second, starting from similar initial states, we carry out long time evolution simulations to quantitatively study the spin relaxation time $T_1$.  The simulation time $t$ is tested and optimized for all the considered magnetic field values and spin bath concentrations.  Close to the LACs, we use 0.1~ms,  while far away from the LACs, we use up to 1~ms simulation time.  Similarly to the simulation time, the time step $dt$ of the time propagation is also optimized.  As a general rule,  $dt$ is selected in a way that even the fastest coherent oscillation is well-resolved in the simulations.  In a sufficiently large spin bath the initial population of the spin states relaxes exponentially,  from which the time scale of the decay ($T_1$) can be obtained. Here, we note that the simulation time is often much shorter than the $T_1$ time.  Due to the extrapolation,  the uncertainty of the calculated $T_1$ time is expectedly larger for a weakly coupled spin bath, where the decay time may reach  seconds.   Ensemble spin relaxation times are obtained by averaging the time-dependent population data over the random spin bath configurations and then fitting an exponential decay curve to the obtained ensemble-averaged population data.  

For high spin defects, the dipolar spin  relaxation depends on the initial spin state.\cite{ramsay_relaxation_2020} For the silicon vacancy, we calculate the spin relaxation effect starting from two different spin states.  The initial population is either evenly distributed in the $m_{\text{S}} = \left\lbrace -1/2, +1/2\right\rbrace$ subspace or set completely in the   $m_{\text{S}} =-1/2$ state.  Since the quartet electron spin is polarized in the $m_{\text{S}} = \left\lbrace -1/2, +1/2\right\rbrace$ subspace in the optical excitation cycle,  the former initial state is natural for microwave-free applications.  High fidelity initialization in a selected spin state can be achieved by applying a resonant microwave pulse.\cite{nagy_high-fidelity_2019}  In all cases, the initial state of the bath spins is set to a thermal state.

\section{Results}
\label{sec:res}

\begin{figure}[h!]
\includegraphics[width=0.525\columnwidth]{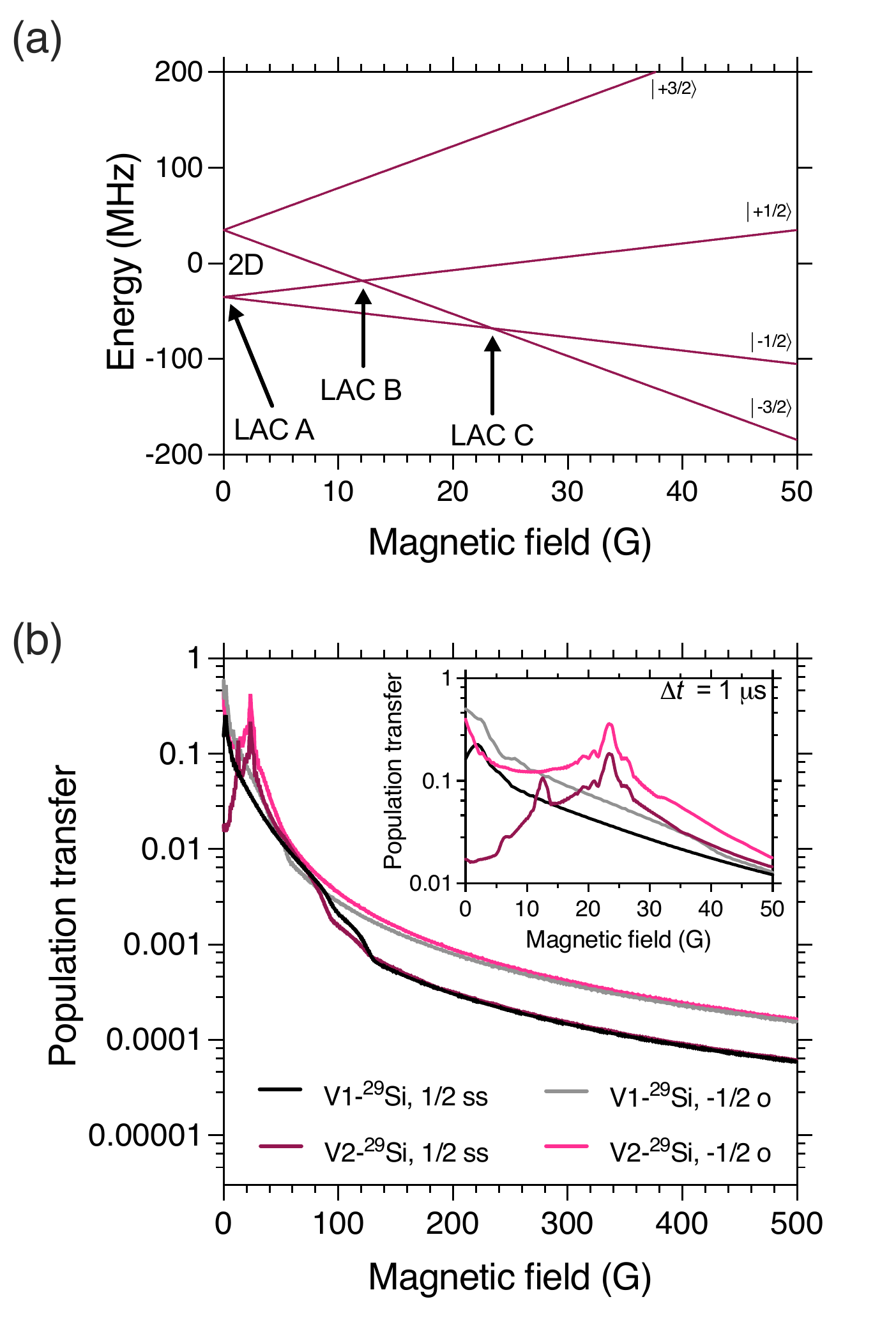}
\caption{  Energy levels of the V2 center and nuclear spin bath induced population transfer between the electron spin states of the silicon vacancy. (a) Spin energy levels of an isolated V2 center in 4H-SiC as a function of the external magnetic field.  Due to the zero-field splitting ($2D$) and the Zeeman shift of the states,  three crossings can be observed,  labelled as LAC~A,  LAC~B, and LAC~C.  The hyperfine interaction with nearby nuclear spins opens a gap between the crossing states and gives rise to LACs .  LACs enable fast spin flip-flops at the corresponding  magnetic field values. (b) Population transfer from the initial spin state to the rest of the silicon vacancy electron spin states.  The inset in (b) depicts the close-up of the zero magnetic field region.   The electron spin is initialized either in the $m_{\text{S}} = \left\lbrace +1/2, -1/2  \right\rbrace$ subspace with equal population (labelled as "1/2 ss") or solely in the  $m_{\text{S}} =-1/2$ spin state (labelled as "-1/2 o").  The LACs induce efficient spin relaxation that rapidly weakens with increasing magnetic field beyond the positions of the LAC~C.  The depicted population transfer curves are obtained after a 1~$\mu$s long time evolution of our model system. 
\label{fig:spin-12n}  }
\end{figure}

First,  we report on spin relaxation effects caused by the surrounding nuclear spin bath of the V1 and V2 centers in 4H-SiC.  In order to understand the results of this section,  we briefly discuss the magnetic field dependence of the quartet energy levels beforehand.  As shown in Fig.~\ref{fig:spin-12n}(a) for the V2 center,  the doubly degenerate $m_{\text{S}} = \left\lbrace -1/2, +1/2\right\rbrace$ and $m_{\text{S}} = \left\lbrace -3/2, 3/2\right\rbrace$ subspaces are split by $2D$ at $B = 0$, due to the ZFS interaction of the $C_{3v}$ symmetric quartet ground state.  The magnetic field applied along the three-fold rotation axis of the defect lifts the initial degeneracies and gives rise to nearly linear Zeeman shifts.  For a positive magnetic field,  the $m_{\text{S}} = -3/2$ level crosses both the $m_{\text{S}} = +1/2$ and the $m_{\text{S}} = -1/2$ levels.  Inclusion of a weakly coupled spin-1/2 nuclear spin does not undermine the depicted level structure, but most importantly its hyperfine interaction gives rise to  LACs at the crossings of electronic spin states of $\Delta m_{\text{S}} = \pm 1$.  The positions of these LACs are labelled by LAC~A and LAC~C in Fig.~\ref{fig:spin-12n}(a).  Furthermore,  second order effects due to couplings that include two nuclear spins,  or a nuclear spin and the non-linear Zeeman terms enable quantum jumps of $\Delta m_{\text{S}} = \pm 2$.  Such second order effects  give rise to an additional LAC midway between LAC~A and LAC~C, which is labelled as LAC~B in Fig.~\ref{fig:spin-12n}(a). At all of these LACs,  enhanced electron spin relaxation is expected due to the  nuclear spin couplings.

In order to qualitatively study the magnetic field dependence of nuclear spin bath induced spin relaxation,  we carry out time evolution simulations up to a fixed time and investigate the amount of population transferred from the initial state to the rest of the electron spin states.  Fig.~\ref{fig:spin-12n}(b) depicts the magnetic field dependence of the obtained population variations for the V1 and the V2 centers for two different initialization conditions denoted by ``1/2 ss'' and ``-1/2 o''.  The former means initialization in the $m_{\text{S}} = \left\lbrace -1/2, +1/2\right\rbrace$ subspace while the latter means initialization in the $m_{\text{S}} =  -1/2$ state only.   In all cases, the population lost from the initial state is polynomially reduced as the magnetic field exceeds $B_{\text{LAC C}}$.  Note that different initialization conditions give rise to different relaxation pathways.  In fact,  since initialization in the $m_{\text{S}} =  -1/2$ enables relaxation to the  $m_{\text{S}} = +1/2 $ state also,  at larger magnetic fields the corresponding population transfer is twice as large as for 1/2~ss subspace initialization. 

For low magnetic field values, the decay of the initial population is sizeable due to the presence of several LACs in the fine energy level structure in this region.  In case of the V2 center, whose ZFS parameter is $D=35$~MHz, all three LACs can be identified in the population transfer plot in Fig.~\ref{fig:spin-12n}(b). Note that LAC~A is observed exclusively for the $m_{\text{S}} =  -1/2$ initialization condition, while LAC~B is observed exclusively when the  $m_{\text{S}} =  +1/2$ state is also populated.   Since the ZFS is $2D = 5.2$~MHz for the V1 center, the LACs can be found very close to $B=0$ for this center.  Due to the width of the LACs, the individual peaks in the  population transition curve cannot be resolved.

\begin{figure}[h!]
\includegraphics[width=0.70\columnwidth]{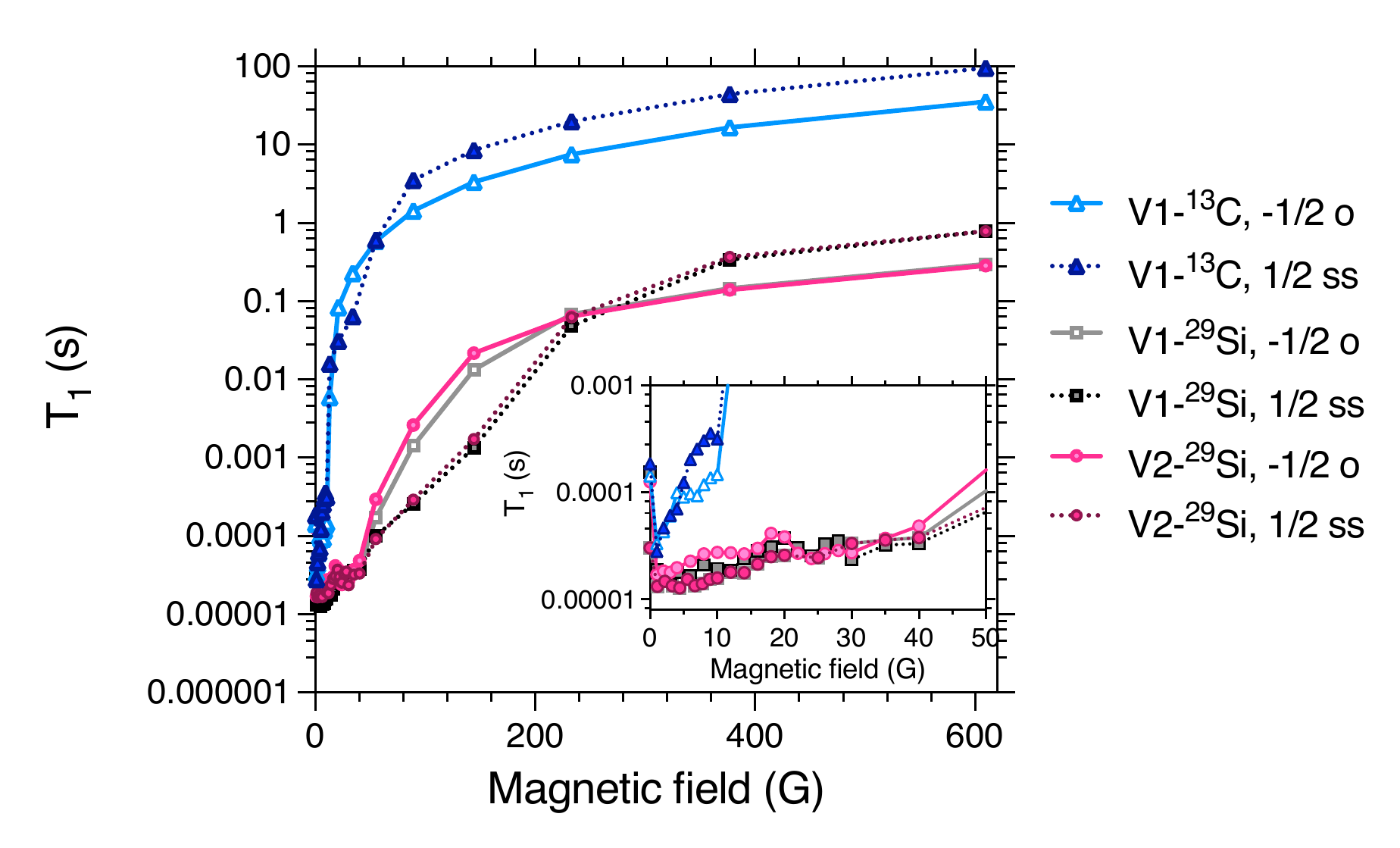}
\caption{  Magnetic field dependence of the ensemble averaged spin relaxation time T$_1$ of the V1 and V2 centers in 4H-SiC  due to hyperfine coupling.  The $^{13}$C and $^{29}$Si nuclear spin bath induced relaxation are provided separately for the V1 center.  Solid and dotted  lines depict relaxation time curves that correspond to different initialization conditions.  For the $^{13}$C nuclear spin bath, the spin is initialized in the $m_{\text{S}} = -1/2$ state only (labelled as "-1/2 o"),  while for the $^{29}$Si nuclear spin bath the spin is initialized in the $ m_{\text{S}} =  \left\lbrace 1/2, -1/2 \right\rbrace$ subspace with equal population in the two states (labelled as 1/2 ss).  The inset highlights relaxation times near zero magnetic field.
\label{fig:spin-12n-T1}  }
\end{figure}

In order to quantitatively analyze the relaxation mechanism of a silicon vacancy qubit in the nuclear spin bath of natural and $^{29}$Si depleted 4H-SiC,  we carry out large-scale numerical simulations to determine the corresponding $T_1$ times.  Fig.~\ref{fig:spin-12n-T1} summarizes the results of these calculations.  For natural samples of $4.68$\% $^{29}$Si and $1.07$\% $^{13}$C isotope content, the relaxation time is determined by the $^{29}$Si nuclear spin and the contribution of the paramagnetic $^{13}$ spins can be neglected.  Indeed, the latter gives rise to a relaxation time which is two orders of magnitude larger than that of $^{29}$Si,  which is the majority nuclear spin source in the sample.  Consequently,  depletion of the $^{29}$Si isotope may give rise to a substantial increase in the relaxation time, when no other relaxation generators,  such as electron spins and spin-phonon coupling, are present.  

The magnetic field dependence of the $T_1$ time clearly shows that the LACs drastically reduce the lifetime of the quartet spin in the  $B =0-60$~Gauss interval for both the  V1 and the V2 centers.  Within this critical region, the $T_1$ expectedly limits the coherence time, while outside of this region  the $T_1$ times exceeds  1~ms,  and the dipolar spin relaxation  does not limit the coherence time anymore.

\begin{figure}[h!]
\includegraphics[width=0.60\columnwidth]{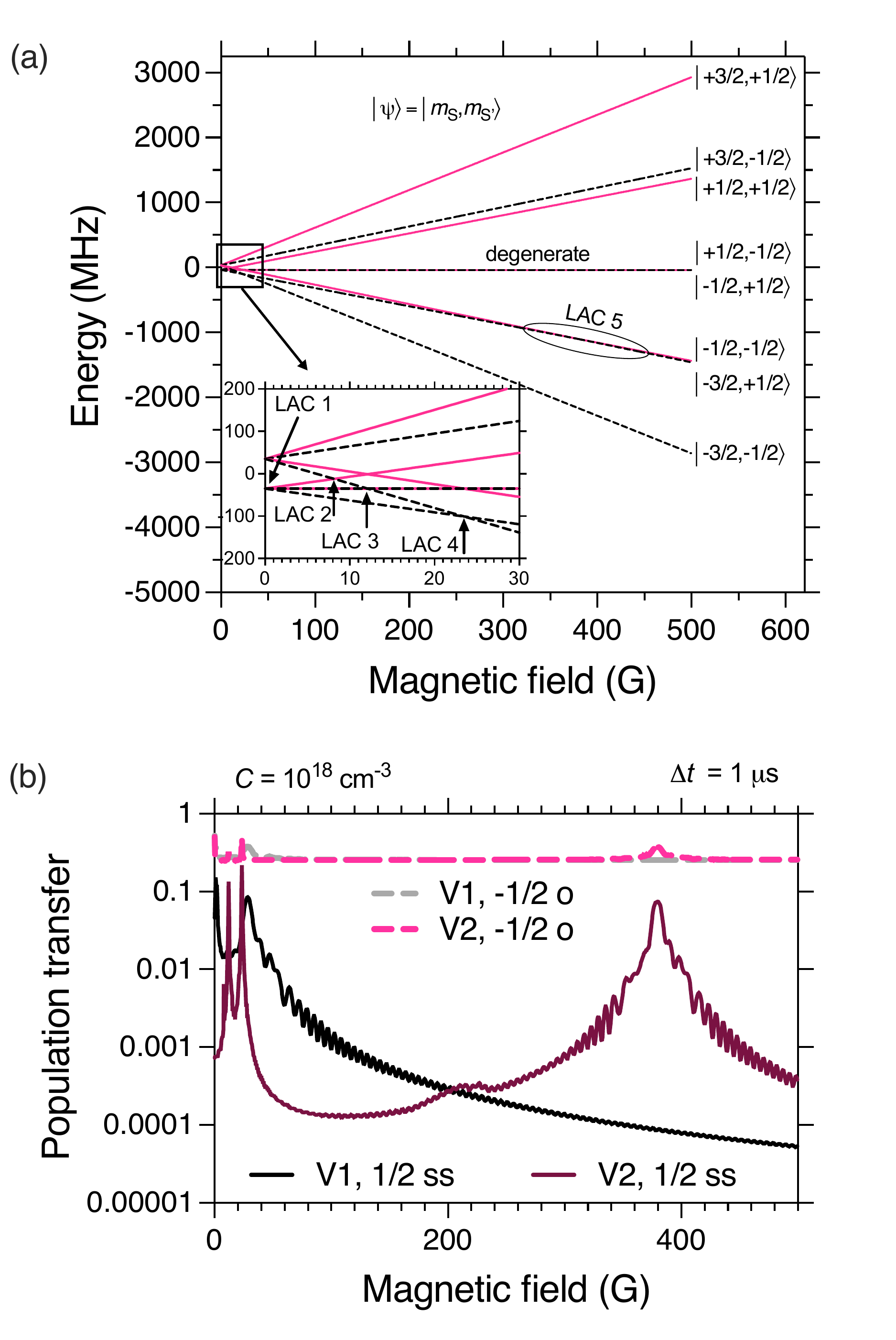}
\caption{  (a) Energy levels of a quartet-doublet two electron spin system.  The inset shows a close-up of the zero magnetic field region.  Altogether five level anti crossings,  marked by LAC~1-5,  can be found at distinct magnetic field values.  The solid pink and the dashed black lines indicates the up and down states of the spin-1/2 defect.  (b) Electron spin defect induced population transfer from the highly polarized initial state to the empty quartet spin states of the V1 and V2 centers.  Solid and dashed lines depict curves that correspond to different initialization conditions.  For the former, the spin is initialized in the $m_{\text{S}} = -1/2$ state only (labelled as "-1/2 o"),  while for the latter the spin is initialized in the $m_{\text{S}} = \left\lbrace 1/2, -1/2 \right\rbrace$ subspace with equal population in the two states (labelled as 1/2 ss). The concentration for the spin-1/2 defect is set to 10$^{18}$~cm$^{-3}$ and an overall simulation time of 1~$\mu$s is used. 
\label{fig:spin-12el}  }
\end{figure}

Next, we investigate spin relaxation effects induced by spin-1/2 electron spins (the case of spin-1 defect environment is briefly discussed in the Appendix).  In SiC, there are several different common paramagnetic defects whose concentration is determined by the growth conditions and after growth treatments.  Ref.~[\onlinecite{bulancea-lindvall_dipolar_2021}] summarizes the most relevant defects and their expected concentrations  in connection with the spin relaxation processes of the divacancy qubits in 4H-SiC.

As depicted in Fig.~\ref{fig:spin-12el}(a),  the fine energy level structure of a quartet-doublet electron spin pair reveals important  differences compared to the energy level structure seen in Fig.~\ref{fig:spin-12n}(a) for the quartet electron spin system.  Since both electron spins exhibit Zeeman splittings with $g \approx 2$ and the zero-field splitting value of the silicon vacancy qubit is already suppressed at small magnetic field values, the magnetic field dependence of the coupled pair's energy levels can be interpreted as the sum of a quintet (spin-2) and a triplet (spin-1) subspaces.  The triplet subspace is nearly degenerate with the three innermost states of the quintet subspace, see Fig.~\ref{fig:spin-12el}(a).  To be able to keep track of the quartet silicon vacancy spin states, hereinafter we use the $\left|  m_{S} m_{S^{\prime}}\right\rangle$ notation to label the states.   Due to the small but non-zero ZFS of the silicon vacancy center, four LACs (LAC1-4) can be found in the vicinity of $B = 0$, see the inset of Fig. ~\ref{fig:spin-12el}(a). Furthermore,  an additional LAC is found at larger magnetic field value (LAC5).  It occurs due to the non-linearity of the quartet spin states of the silicon vacancy.  We note that the position of LAC5 may not be well-defined in a natural sample,  as any inhomogeneity,  for example due to the local hyperfine fields of the two electron spins, may significantly shift the position of LAC5.

Our qualitative study for spin-1/2 electron spin bath induced spin relaxation is summarized in Fig.~\ref{fig:spin-12el}(b).  When the silicon vacancy spin is initialized in the $m_{\text{S}} = \left\lbrace 1/2, -1/2 \right\rbrace$ subspace, three narrow resonance peaks correspond to LAC 2-4, while a wide peak at around $B=380$~Gauss marks the position of LAC 5.  The oscillations observable on the sides of the resonance peak of LAC~5 is a side effect of the finite simulation time and the slow coherent oscillations between states $\left| -3/2,+1/2 \right\rangle$ and  $\left| -1/2,-1/2 \right\rangle$.  For the V1 center, all LAC related resonances occur much closer to $B=0$.

When the initial population is set in the $m_{S} = -1/2$ state only,  we observe an unexpectedly high, nearly magnetic field independent population transfer to the empty states,  see the dashed lines in Fig.~\ref{fig:spin-12el}(b).  To understand this observation, we refer to Fig.~\ref{fig:spin-12el}(a) which shows a degenerate pair of states at the zero value of the energy scale.  This magnetic field independent degenerate subspace includes the $\left| +1/2, -1/2 \right\rangle$ and the $\left| -1/2, +1/2 \right\rangle$  states that can be coupled by the spin flip-flop operator of the dipole-dipole interaction.  Since the states are degenerate,  even a weak coupling between the two electron spins can give rise to a sizeable population transfer within the doublet manifold,  explaining the constant  high value in  Fig.~\ref{fig:spin-12el}(b).  We note that recent measurements have demonstrated fast $\left| -1/2 \right\rangle \leftrightarrow \left| +1/2 \right\rangle $  relaxation\cite{ramsay_relaxation_2020} in accordance with our findings.  The phenomena of efficient spin relaxation between the  $\left| \pm 1/2 \right\rangle $ states  and its effects on the decoherence of the V2 center is further investigated in Ref.~[\onlinecite{oscar_VSi_2}].

\begin{figure}[h!]
\includegraphics[width=0.95\columnwidth]{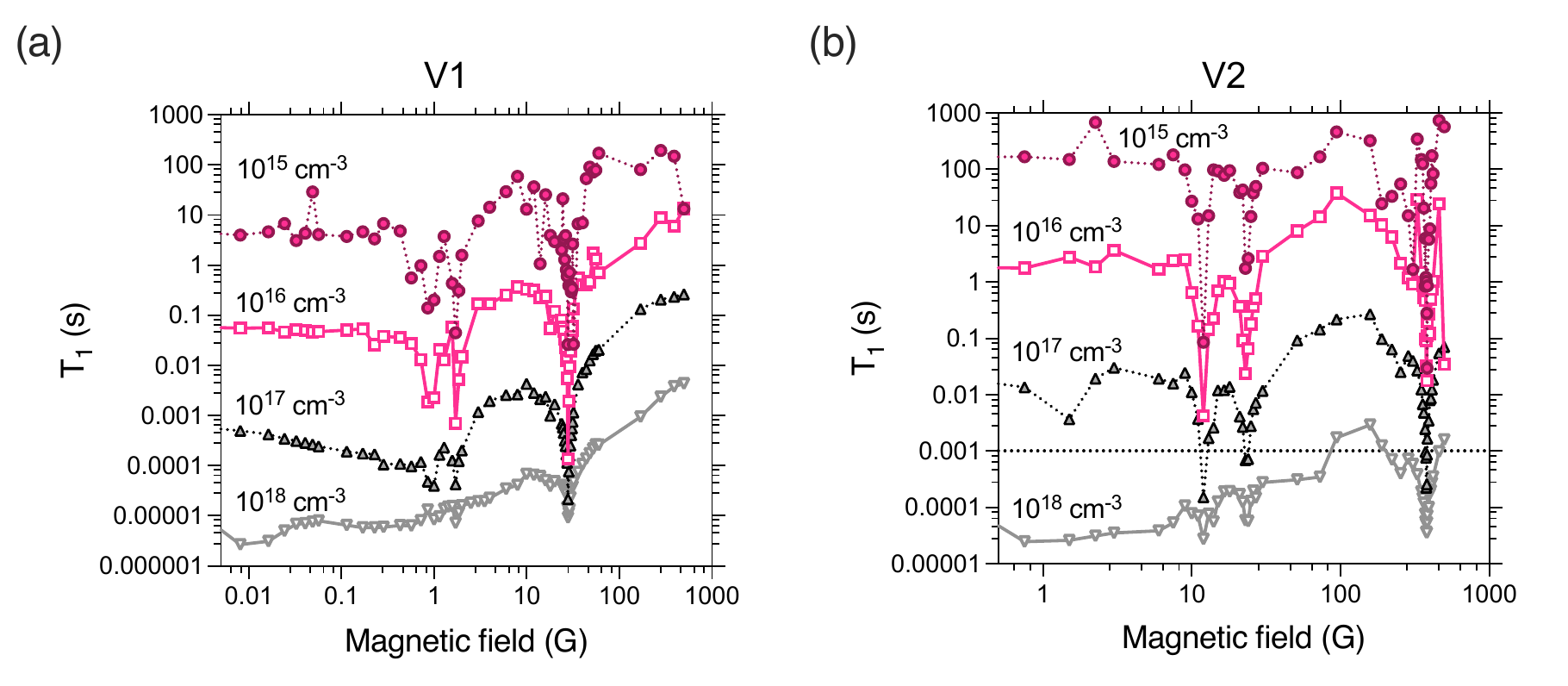}
\caption{  Magnetic field and electron spin defect concentration  dependence of the ensemble averaged spin relaxation time T$_1$ of (a) the V1 center and (b) the V2 center.   The quartet spin state is initialized in the $m_{\text{S}} = \left\lbrace 1/2, -1/2 \right\rbrace$ subspace.  Dashed horizontal line in (b) highlights the typical coherence time for the V2 center.
\label{fig:spin-12-e-T1}  }
\end{figure}

Next, we quantitatively investigate the magnetic field dependence of the longitudinal spin relaxation time of quartet silicon vacancies surrounded by spin-1/2 paramagnetic defects.  Fig.~\ref{fig:spin-12-e-T1} presents the calculated relaxation time of the V1 and the V2 centers for various spin bath concentrations.  As can be seen, the relaxation time ranges from 100~s to 20~$\mu$s depending on the concentration of the defects and the magnetic field.   As expected, the LACs drastically shorten the lifetime of the silicon vacancy spins, however,  the $T_1$ time increases rapidly beyond LAC ~5.

\section{Discussion}
\label{sec:disc}

The careful examination of the most relevant spin relaxation mechanisms of the silicon vacancy qubit presented in section~\ref{sec:res} allows us to propose a low magnetic field relaxometry application,  relying on the variation of the spin state lifetime of the silicon vacancy center in SiC measured through its fluorescence intensity.  Such a microwave-free sensor is desirable for various biological applications.  Furthermore,  the V1 and V2 silicon vacancy centers with 862~nm and 917~nm zero-phonon photo luminescence emit in the first near infra-red transmission window of biological tissue (650~nm and 950~nm),  and thus such a fluorescence silicon vacancy sensor would also be suitable for in vivo applications.

 Since the photoluminescence signal of the silicon vacancy depends on the population of the spin states, spin relaxation-induced population transfers have a direct signature in the optical signal.   Under continuous optical excitation,  the silicon vacancies are probabilistically   excited and the time $t$ spent in the ground state between two excitations follows an exponential distribution.  When the center is in the ground state, the initial high degree of polarization in the $m_{\text{S}} = \pm 1/2$ state exponentially decays.   Therefore,  the PL signal of an ensemble of silicon vacancy centers is equal to
 \begin{equation} \label{eq:signal}
\mathcal{S} = \mathcal{C} \mathcal{I}_0 \int_0^{\infty}  \varrho_{T_d}\! \left( t \right)   e^{-t/
 T_1} dt 
 \end{equation}
 where $ \varrho_{T_d}\! \left( t \right) $ is the probability density function of exponential distribution,  $T_d$ is the average dwell time in the ground state  for a given excitation power,  $\mathcal{C}$ is the spin-dependent contrast of the optical signal,  and $\mathcal{I}_0 $ is the fluorescence intensity of the defect.   Here we assume that the initialization of the spin state through the optical cycle requires negligible time in comparison with  $T_d$.  After integration,  the continuous wave signal is equal to
 \begin{equation}
 \mathcal{S} = \mathcal{C}\mathcal{I}_0 \frac{T_1 }{T_1  + T_d} \text{,}
\end{equation}  
and hence its derivative with respect to the spin relaxation time is
 \begin{equation}
\frac{d\mathcal{S}}{dT_1} = \mathcal{C} \mathcal{I}_0 \left(  - \frac{T_1 }{ \left( T_1  + T_d \right)^2 }  +  \frac{1 }{  T_1  + T_d  } \right) \text{,}
\end{equation}  
 which takes its maximal value at $T_1 / T_d \rightarrow 0$.  It is therefore advisable to set the parameters of silicon vacancy sensors,  excitation laser power and spin defect concentration,  in such a way that  the $T_1 < T_d$ relation is ensured.

Magnetic field sensing can be realized utilizing the magnetic field dependence of the spin relaxation time  $T_1\! \left(B \right)$ studied in this paper.   It is important to note,  however, that the  magnetic field independent spin-orbit and electron-phonon interactions induced spin lattice relaxation $T_1\! \left(\mathcal{T} \right)$, where $\mathcal{T}$ is the temperature, need to be taken into consideration as well, especially at room temperature. The net relaxation rate can be written as a sum of the two terms,
\begin{equation}
\frac{1}{T_1\! \left(B, \mathcal{T} \right)} = \frac{1}{T_1\! \left(\mathcal{T} \right)}  + \frac{1}{T_1\! \left(B\right)} \text{.}
\end{equation}
Based on recent low magnetic field  measurements on the silicon vacancy center in SiC,   $T_1\! \left(\mathcal{T} \right)$  is in the range of  100~$\mu$s at room temperature. \cite{simin_locking_2017,singh_experimental_2020} In order to obtain a measurable magnetic field dependent signal, the dipole-dipole interaction induced spin relaxation time $T_1\! \left(B\right)$ needs to be at least in the same order of magnitude.  For the V1 (V2) center this can be achieved at low magnetic field strength by using  $C = 10^{17}$~cm$^{-3}$ ($C = 3 \times 10^{17}$~cm$^{-3}$ ) spin-1/2 defect concentration in the host material. 

In order to achieve the highest sensitivity,  the derivative of the relaxation time with respect to the magnetic field $ dT_1\! \left(B\right) / dB $  should be maximal.  In this respect, the hyperfine interaction that gives rise to local inhomogeneities, variation of the LAC positions, and broadening of the resonance signal is undesirable.  Therefore,  for high performance sensing applications $^{29}$Si (and $^{13}C$) depleted samples are needed.  In our calculations, we consider relaxation mechanisms in nuclear and electron spin bath separately,  thus our  results for spin-1/2 electron spin environments  correspond to isotope purified samples.  

\begin{figure}[h!]
\includegraphics[width=0.8\columnwidth]{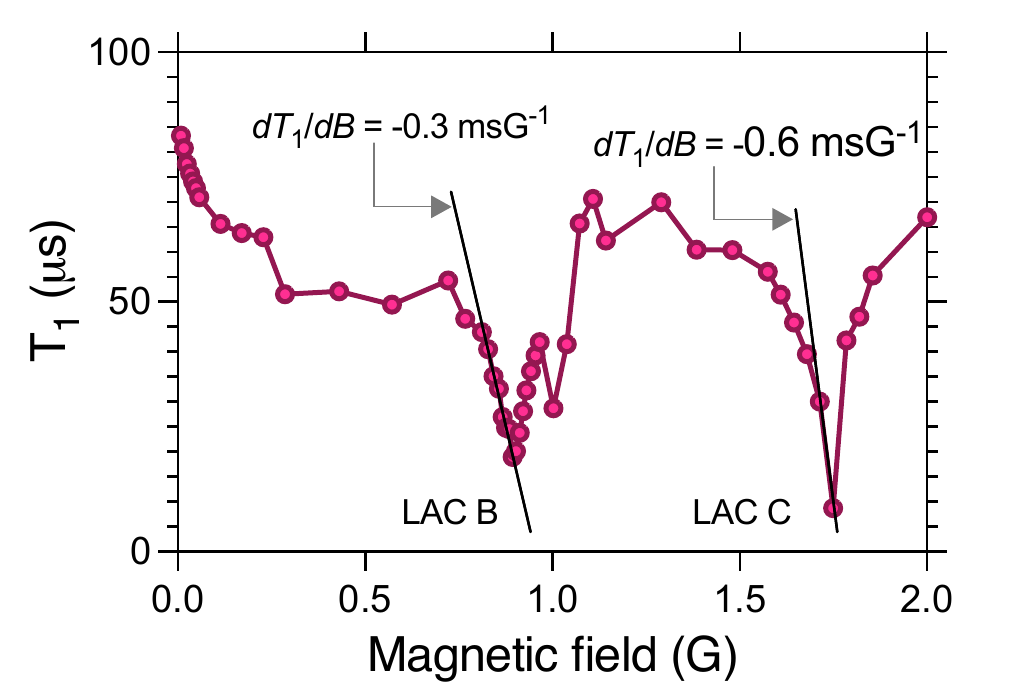}
\caption{  Spin relaxation curve of the V1 center at room temperature ($T_1 \! \left( 300~K\right) = 100$~$\mu$s) for $C = 10^{17}$~cm$^{-3}$ spin-1/2 point defect concentration in isotope purified 4H-SiC sample.
\label{fig:grad}  }
\end{figure}

The highest gradients can be found close to zero magnetic field at the  sharp resonances of the spin relaxation curves that are associated with LACs in the energy level structure.  In Fig.~\ref{fig:grad}, we depict the $T_1 \! \left( B, 300\text{ K} \right)$ curve of the V1 center close to zero magnetic field for $C = 10^{17}$~cm$^{-3}$ spin-1/2 defect concentration in paramagnetic isotope purified 4H-SiC sample.  The largest derivative of 0.6~msGauss$^{-1}$ is obtained at an external magnetic field value of 1.75~G.   Inserting this gradient into Eq.~(\ref{eq:signal}) we obtain the variation of the PL signal through the following formula
\begin{equation}
\frac{1}{\mathcal{C} \mathcal{I}_0} \left|  \frac{ d\mathcal{S} }{dB} \right|  = 4.2 \text{ G}^{-1} \text{.}
\end{equation}
This implies that the PL signal changes by 4.2\%  of the spin contrast $\mathcal{C}$ for $\Delta B = 1$~$\mu$T at $B = 0.175$~mT external field.  Such a high gradient may lead to the realization of sensitive DC magnetic field sensors in SiC at very small magnetic field values.

\section{Summary}
\label{sec:sum}

In summary,  we comprehensively investigated the dipolar spin relaxation of quartet silicon vacancy qubits in 4H-SiC.  The observed relaxation phenomena significantly  differ  from the case of triplet qubits in wide band gap semiconductors,  such as the NV-center in diamond and the divacancy in SiC.  Due to the small zero-field splitting, the LACs and most of the strong relaxation processes squeeze into a small magnetic field interval  close to $B=0$.  While this region is generally avoided in experiments due to the strong couplings,  here we show that one may utilize these effects in  room-temperature microwave-free magnetic field sensing applications.  Based on our results we propose that the most suitable 4H-SiC sample for this purpuse is a paramagnetic isotope purified sample that contains spin-1/2 defects in $\approx 10^{17}$~cm$^{-3}$ concentration.

\section*{Acknowledgments}

We acknowledge support from the Knut and Alice Wallenberg Foundation through WBSQD2 project (Grant No.\ KAW 2018.0071). Support from the Swedish Government Strategic Research Area SeRC and the Swedish Government Strategic Research Area in Materials Science on Functional Materials at Link\"{o}ping University (Faculty Grant SFO-Mat-LiU No. 2009 00971) is gratefully acknowledged.  VI acknowledges the support from the MTA Premium Postdoctoral Research Program. N. T. S. acknowledges the support from the Swedish Research Council (Grant No. VR 2016-04068), the EU H2020 project QuanTELCO (Grant No. 862721). The calculations were performed on resources provided by the Swedish National Infrastructure for Computing (SNIC) and Liu local at the National Supercomputer Centre (NSC).

\section*{Appendix: Spin relaxation due to spin-1 defects} 

\begin{figure}[h!]
\includegraphics[width=0.8\columnwidth]{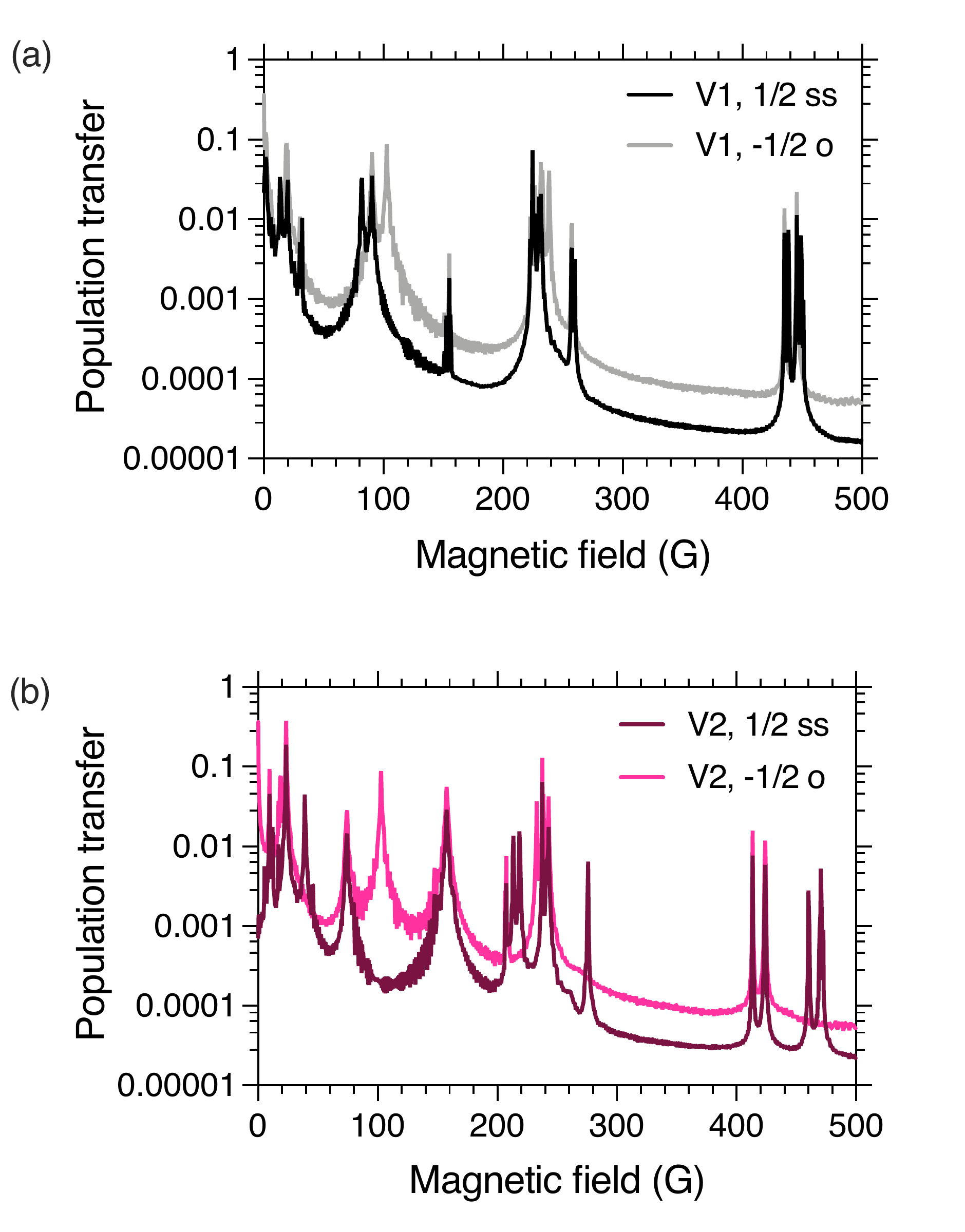}
\caption{  Population transfer between a silicon vacancy and a bath of  divacancy qubits in 4H-SiC.  (a) and (b) depict the case of V1 and V2 centers, respectively.  Dark and light color curves show different spin initialization conditions.  In both cases the figure integrates the contributions from all of the four different divacancy configurations.
\label{fig:spin-1}  }
\end{figure}

In this appendix, we qualitatively investigate the coupling of the quartet silicon vacancy spin states to a bath of divacancies.  The concentration  of these spin-1 defects is not substantial in commonly studied 4H-SiC samples,  thus their contribution to the spin relaxation time may be negligible in most cases.  On the other hand,  when the electron spin states are resonant, even a few neighbouring divacancy centers can give rise to sizeable relaxation effects that may be undesirable from the application point of view.  To reveal the magnetic field values of the enhanced relaxation due to divacancy  spins, we study spin bath coupling induced polarization transfer between the initially polarized and the initially empty states over a 1~$\mu$s simulation time, see Figs.~\ref{fig:spin-1}(a) and (b) for the V1 and V2 centers.   Note that the spin bath includes all four possible divacancy configurations, thus Fig.~\ref{fig:spin-1} shows the integrated effect of all the different divacancies.  As can be seen in Fig.~\ref{fig:spin-1}, there are numerous resonances due to the multiple crossings between electronic states and the variance of the $D_i$ and $E_i$ ZFS parameters. Furthermore,  the relative amplitude of certain resonance peaks may vary depending on the initial spin state of the silicon vacancy and divacancy qubits.

As can be seen in Fig.~\ref{fig:spin-1},  the number of resonance peaks drops as the magnetic field increases. In particular,  for 300~G $< B <$ 400~G and for 500~G $< B$ no resonance peaks can be found. These magnetic field regions may be advantageous for silicon vacancy applications when both silicon vacancies and divacancy defects are created by positive ion implantation and subsequent annealing.


\begin{thebibliography}{40}%
\makeatletter
\providecommand \@ifxundefined [1]{%
 \@ifx{#1\undefined}
}%
\providecommand \@ifnum [1]{%
 \ifnum #1\expandafter \@firstoftwo
 \else \expandafter \@secondoftwo
 \fi
}%
\providecommand \@ifx [1]{%
 \ifx #1\expandafter \@firstoftwo
 \else \expandafter \@secondoftwo
 \fi
}%
\providecommand \natexlab [1]{#1}%
\providecommand \enquote  [1]{``#1''}%
\providecommand \bibnamefont  [1]{#1}%
\providecommand \bibfnamefont [1]{#1}%
\providecommand \citenamefont [1]{#1}%
\providecommand \href@noop [0]{\@secondoftwo}%
\providecommand \href [0]{\begingroup \@sanitize@url \@href}%
\providecommand \@href[1]{\@@startlink{#1}\@@href}%
\providecommand \@@href[1]{\endgroup#1\@@endlink}%
\providecommand \@sanitize@url [0]{\catcode `\\12\catcode `\$12\catcode
  `\&12\catcode `\#12\catcode `\^12\catcode `\_12\catcode `\%12\relax}%
\providecommand \@@startlink[1]{}%
\providecommand \@@endlink[0]{}%
\providecommand \url  [0]{\begingroup\@sanitize@url \@url }%
\providecommand \@url [1]{\endgroup\@href {#1}{\urlprefix }}%
\providecommand \urlprefix  [0]{URL }%
\providecommand \Eprint [0]{\href }%
\providecommand \doibase [0]{http://dx.doi.org/}%
\providecommand \selectlanguage [0]{\@gobble}%
\providecommand \bibinfo  [0]{\@secondoftwo}%
\providecommand \bibfield  [0]{\@secondoftwo}%
\providecommand \translation [1]{[#1]}%
\providecommand \BibitemOpen [0]{}%
\providecommand \bibitemStop [0]{}%
\providecommand \bibitemNoStop [0]{.\EOS\space}%
\providecommand \EOS [0]{\spacefactor3000\relax}%
\providecommand \BibitemShut  [1]{\csname bibitem#1\endcsname}%
\let\auto@bib@innerbib\@empty
\bibitem [{\citenamefont {Doherty}\ \emph {et~al.}(2013)\citenamefont
  {Doherty}, \citenamefont {Manson}, \citenamefont {Delaney}, \citenamefont
  {Jelezko}, \citenamefont {Wrachtrup},\ and\ \citenamefont
  {Hollenberg}}]{DohertyNVreview}%
  \BibitemOpen
  \bibfield  {author} {\bibinfo {author} {\bibfnamefont {M.~W.}\ \bibnamefont
  {Doherty}}, \bibinfo {author} {\bibfnamefont {N.~B.}\ \bibnamefont {Manson}},
  \bibinfo {author} {\bibfnamefont {P.}~\bibnamefont {Delaney}}, \bibinfo
  {author} {\bibfnamefont {F.}~\bibnamefont {Jelezko}}, \bibinfo {author}
  {\bibfnamefont {J.}~\bibnamefont {Wrachtrup}}, \ and\ \bibinfo {author}
  {\bibfnamefont {L.~C.}\ \bibnamefont {Hollenberg}},\ }\href {\doibase
  https://doi.org/10.1016/j.physrep.2013.02.001} {\bibfield  {journal}
  {\bibinfo  {journal} {Physics Reports}\ }\textbf {\bibinfo {volume} {528}},\
  \bibinfo {pages} {1 } (\bibinfo {year} {2013})}\BibitemShut {NoStop}%
\bibitem [{\citenamefont {Barry}\ \emph {et~al.}(2020)\citenamefont {Barry},
  \citenamefont {Schloss}, \citenamefont {Bauch}, \citenamefont {Turner},
  \citenamefont {Hart}, \citenamefont {Pham},\ and\ \citenamefont
  {Walsworth}}]{barry_sensitivity_2020}%
  \BibitemOpen
  \bibfield  {author} {\bibinfo {author} {\bibfnamefont {J.~F.}\ \bibnamefont
  {Barry}}, \bibinfo {author} {\bibfnamefont {J.~M.}\ \bibnamefont {Schloss}},
  \bibinfo {author} {\bibfnamefont {E.}~\bibnamefont {Bauch}}, \bibinfo
  {author} {\bibfnamefont {M.~J.}\ \bibnamefont {Turner}}, \bibinfo {author}
  {\bibfnamefont {C.~A.}\ \bibnamefont {Hart}}, \bibinfo {author}
  {\bibfnamefont {L.~M.}\ \bibnamefont {Pham}}, \ and\ \bibinfo {author}
  {\bibfnamefont {R.~L.}\ \bibnamefont {Walsworth}},\ }\href {\doibase
  10.1103/RevModPhys.92.015004} {\bibfield  {journal} {\bibinfo  {journal}
  {Reviews of Modern Physics}\ }\textbf {\bibinfo {volume} {92}},\ \bibinfo
  {pages} {015004} (\bibinfo {year} {2020})}\BibitemShut {NoStop}%
\bibitem [{\citenamefont {Degen}\ \emph {et~al.}(2017)\citenamefont {Degen},
  \citenamefont {Reinhard},\ and\ \citenamefont
  {Cappellaro}}]{degen_quantum_2017}%
  \BibitemOpen
  \bibfield  {author} {\bibinfo {author} {\bibfnamefont {C.}~\bibnamefont
  {Degen}}, \bibinfo {author} {\bibfnamefont {F.}~\bibnamefont {Reinhard}}, \
  and\ \bibinfo {author} {\bibfnamefont {P.}~\bibnamefont {Cappellaro}},\
  }\href {\doibase 10.1103/RevModPhys.89.035002} {\bibfield  {journal}
  {\bibinfo  {journal} {Reviews of Modern Physics}\ }\textbf {\bibinfo {volume}
  {89}},\ \bibinfo {pages} {035002} (\bibinfo {year} {2017})}\BibitemShut
  {NoStop}%
\bibitem [{\citenamefont {Tetienne}\ \emph {et~al.}(2013)\citenamefont
  {Tetienne}, \citenamefont {Hingant}, \citenamefont {Rondin}, \citenamefont
  {Cavaill\`es}, \citenamefont {Mayer}, \citenamefont {Dantelle}, \citenamefont
  {Gacoin}, \citenamefont {Wrachtrup}, \citenamefont {Roch},\ and\
  \citenamefont {Jacques}}]{tetienne_spin_2013}%
  \BibitemOpen
  \bibfield  {author} {\bibinfo {author} {\bibfnamefont {J.-P.}\ \bibnamefont
  {Tetienne}}, \bibinfo {author} {\bibfnamefont {T.}~\bibnamefont {Hingant}},
  \bibinfo {author} {\bibfnamefont {L.}~\bibnamefont {Rondin}}, \bibinfo
  {author} {\bibfnamefont {A.}~\bibnamefont {Cavaill\`es}}, \bibinfo {author}
  {\bibfnamefont {L.}~\bibnamefont {Mayer}}, \bibinfo {author} {\bibfnamefont
  {G.}~\bibnamefont {Dantelle}}, \bibinfo {author} {\bibfnamefont
  {T.}~\bibnamefont {Gacoin}}, \bibinfo {author} {\bibfnamefont
  {J.}~\bibnamefont {Wrachtrup}}, \bibinfo {author} {\bibfnamefont {J.-F.}\
  \bibnamefont {Roch}}, \ and\ \bibinfo {author} {\bibfnamefont
  {V.}~\bibnamefont {Jacques}},\ }\href {\doibase 10.1103/PhysRevB.87.235436}
  {\bibfield  {journal} {\bibinfo  {journal} {Phys. Rev. B}\ }\textbf {\bibinfo
  {volume} {87}},\ \bibinfo {pages} {235436} (\bibinfo {year}
  {2013})}\BibitemShut {NoStop}%
\bibitem [{\citenamefont {Schmid-Lorch}\ \emph {et~al.}(2015)\citenamefont
  {Schmid-Lorch}, \citenamefont {Häberle}, \citenamefont {Reinhard},
  \citenamefont {Zappe}, \citenamefont {Slota}, \citenamefont {Bogani},
  \citenamefont {Finkler},\ and\ \citenamefont
  {Wrachtrup}}]{schmid-lorch_relaxometry_2015}%
  \BibitemOpen
  \bibfield  {author} {\bibinfo {author} {\bibfnamefont {D.}~\bibnamefont
  {Schmid-Lorch}}, \bibinfo {author} {\bibfnamefont {T.}~\bibnamefont
  {Häberle}}, \bibinfo {author} {\bibfnamefont {F.}~\bibnamefont {Reinhard}},
  \bibinfo {author} {\bibfnamefont {A.}~\bibnamefont {Zappe}}, \bibinfo
  {author} {\bibfnamefont {M.}~\bibnamefont {Slota}}, \bibinfo {author}
  {\bibfnamefont {L.}~\bibnamefont {Bogani}}, \bibinfo {author} {\bibfnamefont
  {A.}~\bibnamefont {Finkler}}, \ and\ \bibinfo {author} {\bibfnamefont
  {J.}~\bibnamefont {Wrachtrup}},\ }\href {\doibase
  10.1021/acs.nanolett.5b00679} {\bibfield  {journal} {\bibinfo  {journal}
  {Nano Letters}\ }\textbf {\bibinfo {volume} {15}},\ \bibinfo {pages} {4942}
  (\bibinfo {year} {2015})}\BibitemShut {NoStop}%
\bibitem [{\citenamefont {van~der Sar}\ \emph {et~al.}(2015)\citenamefont
  {van~der Sar}, \citenamefont {Casola}, \citenamefont {Walsworth},\ and\
  \citenamefont {Yacoby}}]{van_der_sar_nanometre-scale_2015}%
  \BibitemOpen
  \bibfield  {author} {\bibinfo {author} {\bibfnamefont {T.}~\bibnamefont
  {van~der Sar}}, \bibinfo {author} {\bibfnamefont {F.}~\bibnamefont {Casola}},
  \bibinfo {author} {\bibfnamefont {R.}~\bibnamefont {Walsworth}}, \ and\
  \bibinfo {author} {\bibfnamefont {A.}~\bibnamefont {Yacoby}},\ }\href
  {\doibase 10.1038/ncomms8886} {\bibfield  {journal} {\bibinfo  {journal}
  {Nature Communications}\ }\textbf {\bibinfo {volume} {6}},\ \bibinfo {pages}
  {7886} (\bibinfo {year} {2015})}\BibitemShut {NoStop}%
\bibitem [{\citenamefont {Hall}\ \emph {et~al.}(2016)\citenamefont {Hall},
  \citenamefont {Kehayias}, \citenamefont {Simpson}, \citenamefont {Jarmola},
  \citenamefont {Stacey}, \citenamefont {Budker},\ and\ \citenamefont
  {Hollenberg}}]{hall_detection_2016}%
  \BibitemOpen
  \bibfield  {author} {\bibinfo {author} {\bibfnamefont {L.~T.}\ \bibnamefont
  {Hall}}, \bibinfo {author} {\bibfnamefont {P.}~\bibnamefont {Kehayias}},
  \bibinfo {author} {\bibfnamefont {D.~A.}\ \bibnamefont {Simpson}}, \bibinfo
  {author} {\bibfnamefont {A.}~\bibnamefont {Jarmola}}, \bibinfo {author}
  {\bibfnamefont {A.}~\bibnamefont {Stacey}}, \bibinfo {author} {\bibfnamefont
  {D.}~\bibnamefont {Budker}}, \ and\ \bibinfo {author} {\bibfnamefont
  {L.~C.~L.}\ \bibnamefont {Hollenberg}},\ }\href {\doibase
  10.1038/ncomms10211} {\bibfield  {journal} {\bibinfo  {journal} {Nature
  Communications}\ }\textbf {\bibinfo {volume} {7}},\ \bibinfo {pages} {10211}
  (\bibinfo {year} {2016})}\BibitemShut {NoStop}%
\bibitem [{\citenamefont {Rendler}\ \emph {et~al.}(2017)\citenamefont
  {Rendler}, \citenamefont {Neburkova}, \citenamefont {Zemek}, \citenamefont
  {Kotek}, \citenamefont {Zappe}, \citenamefont {Chu}, \citenamefont {Cigler},\
  and\ \citenamefont {Wrachtrup}}]{rendler_optical_2017}%
  \BibitemOpen
  \bibfield  {author} {\bibinfo {author} {\bibfnamefont {T.}~\bibnamefont
  {Rendler}}, \bibinfo {author} {\bibfnamefont {J.}~\bibnamefont {Neburkova}},
  \bibinfo {author} {\bibfnamefont {O.}~\bibnamefont {Zemek}}, \bibinfo
  {author} {\bibfnamefont {J.}~\bibnamefont {Kotek}}, \bibinfo {author}
  {\bibfnamefont {A.}~\bibnamefont {Zappe}}, \bibinfo {author} {\bibfnamefont
  {Z.}~\bibnamefont {Chu}}, \bibinfo {author} {\bibfnamefont {P.}~\bibnamefont
  {Cigler}}, \ and\ \bibinfo {author} {\bibfnamefont {J.}~\bibnamefont
  {Wrachtrup}},\ }\href {\doibase 10.1038/ncomms14701} {\bibfield  {journal}
  {\bibinfo  {journal} {Nature Communications}\ }\textbf {\bibinfo {volume}
  {8}},\ \bibinfo {pages} {14701} (\bibinfo {year} {2017})}\BibitemShut
  {NoStop}%
\bibitem [{\citenamefont {Finco}\ \emph {et~al.}(2021)\citenamefont {Finco},
  \citenamefont {Haykal}, \citenamefont {Tanos}, \citenamefont {Fabre},
  \citenamefont {Chouaieb}, \citenamefont {Akhtar}, \citenamefont
  {Robert-Philip}, \citenamefont {Legrand}, \citenamefont {Ajejas},
  \citenamefont {Bouzehouane}, \citenamefont {Reyren}, \citenamefont
  {Devolder}, \citenamefont {Adam}, \citenamefont {Kim}, \citenamefont {Cros},\
  and\ \citenamefont {Jacques}}]{finco_imaging_2021}%
  \BibitemOpen
  \bibfield  {author} {\bibinfo {author} {\bibfnamefont {A.}~\bibnamefont
  {Finco}}, \bibinfo {author} {\bibfnamefont {A.}~\bibnamefont {Haykal}},
  \bibinfo {author} {\bibfnamefont {R.}~\bibnamefont {Tanos}}, \bibinfo
  {author} {\bibfnamefont {F.}~\bibnamefont {Fabre}}, \bibinfo {author}
  {\bibfnamefont {S.}~\bibnamefont {Chouaieb}}, \bibinfo {author}
  {\bibfnamefont {W.}~\bibnamefont {Akhtar}}, \bibinfo {author} {\bibfnamefont
  {I.}~\bibnamefont {Robert-Philip}}, \bibinfo {author} {\bibfnamefont
  {W.}~\bibnamefont {Legrand}}, \bibinfo {author} {\bibfnamefont
  {F.}~\bibnamefont {Ajejas}}, \bibinfo {author} {\bibfnamefont
  {K.}~\bibnamefont {Bouzehouane}}, \bibinfo {author} {\bibfnamefont
  {N.}~\bibnamefont {Reyren}}, \bibinfo {author} {\bibfnamefont
  {T.}~\bibnamefont {Devolder}}, \bibinfo {author} {\bibfnamefont {J.-P.}\
  \bibnamefont {Adam}}, \bibinfo {author} {\bibfnamefont {J.-V.}\ \bibnamefont
  {Kim}}, \bibinfo {author} {\bibfnamefont {V.}~\bibnamefont {Cros}}, \ and\
  \bibinfo {author} {\bibfnamefont {V.}~\bibnamefont {Jacques}},\ }\href
  {\doibase 10.1038/s41467-021-20995-x} {\bibfield  {journal} {\bibinfo
  {journal} {Nature Communications}\ }\textbf {\bibinfo {volume} {12}},\
  \bibinfo {pages} {767} (\bibinfo {year} {2021})}\BibitemShut {NoStop}%
\bibitem [{\citenamefont {Zheng}\ \emph {et~al.}(2019)\citenamefont {Zheng},
  \citenamefont {Xu}, \citenamefont {Iwata}, \citenamefont {Lenz},
  \citenamefont {Michl}, \citenamefont {Yavkin}, \citenamefont {Nakamura},
  \citenamefont {Sumiya}, \citenamefont {Ohshima}, \citenamefont {Isoya},
  \citenamefont {Wrachtrup}, \citenamefont {Wickenbrock},\ and\ \citenamefont
  {Budker}}]{zheng_zero-field_2019}%
  \BibitemOpen
  \bibfield  {author} {\bibinfo {author} {\bibfnamefont {H.}~\bibnamefont
  {Zheng}}, \bibinfo {author} {\bibfnamefont {J.}~\bibnamefont {Xu}}, \bibinfo
  {author} {\bibfnamefont {G.~Z.}\ \bibnamefont {Iwata}}, \bibinfo {author}
  {\bibfnamefont {T.}~\bibnamefont {Lenz}}, \bibinfo {author} {\bibfnamefont
  {J.}~\bibnamefont {Michl}}, \bibinfo {author} {\bibfnamefont
  {B.}~\bibnamefont {Yavkin}}, \bibinfo {author} {\bibfnamefont
  {K.}~\bibnamefont {Nakamura}}, \bibinfo {author} {\bibfnamefont
  {H.}~\bibnamefont {Sumiya}}, \bibinfo {author} {\bibfnamefont
  {T.}~\bibnamefont {Ohshima}}, \bibinfo {author} {\bibfnamefont
  {J.}~\bibnamefont {Isoya}}, \bibinfo {author} {\bibfnamefont
  {J.}~\bibnamefont {Wrachtrup}}, \bibinfo {author} {\bibfnamefont
  {A.}~\bibnamefont {Wickenbrock}}, \ and\ \bibinfo {author} {\bibfnamefont
  {D.}~\bibnamefont {Budker}},\ }\href {\doibase
  10.1103/PhysRevApplied.11.064068} {\bibfield  {journal} {\bibinfo  {journal}
  {Physical Review Applied}\ }\textbf {\bibinfo {volume} {11}},\ \bibinfo
  {pages} {064068} (\bibinfo {year} {2019})}\BibitemShut {NoStop}%
\bibitem [{\citenamefont {Lenz}\ \emph {et~al.}(2020)\citenamefont {Lenz},
  \citenamefont {Wickenbrock}, \citenamefont {Jelezko}, \citenamefont
  {Balasubramanian},\ and\ \citenamefont {Budker}}]{lenz_magnetic_2020}%
  \BibitemOpen
  \bibfield  {author} {\bibinfo {author} {\bibfnamefont {T.}~\bibnamefont
  {Lenz}}, \bibinfo {author} {\bibfnamefont {A.}~\bibnamefont {Wickenbrock}},
  \bibinfo {author} {\bibfnamefont {F.}~\bibnamefont {Jelezko}}, \bibinfo
  {author} {\bibfnamefont {G.}~\bibnamefont {Balasubramanian}}, \ and\ \bibinfo
  {author} {\bibfnamefont {D.}~\bibnamefont {Budker}},\ }\href
  {http://arxiv.org/abs/2009.12117} {\bibfield  {journal} {\bibinfo  {journal}
  {arXiv:2009.12117 [physics]}\ } (\bibinfo {year} {2020})},\ \bibinfo {note}
  {arXiv: 2009.12117}\BibitemShut {NoStop}%
\bibitem [{\citenamefont {Fu}\ \emph {et~al.}(2020)\citenamefont {Fu},
  \citenamefont {Iwata}, \citenamefont {Wickenbrock},\ and\ \citenamefont
  {Budker}}]{fu_sensitive_2020}%
  \BibitemOpen
  \bibfield  {author} {\bibinfo {author} {\bibfnamefont {K.-M.~C.}\
  \bibnamefont {Fu}}, \bibinfo {author} {\bibfnamefont {G.~Z.}\ \bibnamefont
  {Iwata}}, \bibinfo {author} {\bibfnamefont {A.}~\bibnamefont {Wickenbrock}},
  \ and\ \bibinfo {author} {\bibfnamefont {D.}~\bibnamefont {Budker}},\ }\href
  {\doibase 10.1116/5.0025186} {\bibfield  {journal} {\bibinfo  {journal} {AVS
  Quantum Science}\ }\textbf {\bibinfo {volume} {2}},\ \bibinfo {pages}
  {044702} (\bibinfo {year} {2020})}\BibitemShut {NoStop}%
\bibitem [{\citenamefont {Wang}\ \emph {et~al.}(2021)\citenamefont {Wang},
  \citenamefont {Liu}, \citenamefont {Fan}, \citenamefont {Feng}, \citenamefont
  {Leong}, \citenamefont {Finkler}, \citenamefont {Denisenko}, \citenamefont
  {Wrachtrup}, \citenamefont {Li},\ and\ \citenamefont
  {Liu}}]{wang_zero-field_2021}%
  \BibitemOpen
  \bibfield  {author} {\bibinfo {author} {\bibfnamefont {N.}~\bibnamefont
  {Wang}}, \bibinfo {author} {\bibfnamefont {C.-F.}\ \bibnamefont {Liu}},
  \bibinfo {author} {\bibfnamefont {J.-W.}\ \bibnamefont {Fan}}, \bibinfo
  {author} {\bibfnamefont {X.}~\bibnamefont {Feng}}, \bibinfo {author}
  {\bibfnamefont {W.-H.}\ \bibnamefont {Leong}}, \bibinfo {author}
  {\bibfnamefont {A.}~\bibnamefont {Finkler}}, \bibinfo {author} {\bibfnamefont
  {A.}~\bibnamefont {Denisenko}}, \bibinfo {author} {\bibfnamefont
  {J.}~\bibnamefont {Wrachtrup}}, \bibinfo {author} {\bibfnamefont
  {Q.}~\bibnamefont {Li}}, \ and\ \bibinfo {author} {\bibfnamefont {R.-B.}\
  \bibnamefont {Liu}},\ }\href {http://arxiv.org/abs/2109.05445} {\bibfield
  {journal} {\bibinfo  {journal} {arXiv:2109.05445 [cond-mat,
  physics:quant-ph]}\ } (\bibinfo {year} {2021})},\ \bibinfo {note} {arXiv:
  2109.05445}\BibitemShut {NoStop}%
\bibitem [{\citenamefont {Koehl}\ \emph {et~al.}(2011)\citenamefont {Koehl},
  \citenamefont {Buckley}, \citenamefont {Heremans}, \citenamefont {Calusine},\
  and\ \citenamefont {Awschalom}}]{koehl_room_2011}%
  \BibitemOpen
  \bibfield  {author} {\bibinfo {author} {\bibfnamefont {W.~F.}\ \bibnamefont
  {Koehl}}, \bibinfo {author} {\bibfnamefont {B.~B.}\ \bibnamefont {Buckley}},
  \bibinfo {author} {\bibfnamefont {F.~J.}\ \bibnamefont {Heremans}}, \bibinfo
  {author} {\bibfnamefont {G.}~\bibnamefont {Calusine}}, \ and\ \bibinfo
  {author} {\bibfnamefont {D.~D.}\ \bibnamefont {Awschalom}},\ }\href {\doibase
  10.1038/nature10562} {\bibfield  {journal} {\bibinfo  {journal} {Nature}\
  }\textbf {\bibinfo {volume} {479}},\ \bibinfo {pages} {84} (\bibinfo {year}
  {2011})}\BibitemShut {NoStop}%
\bibitem [{\citenamefont {Bulancea-Lindvall}\ \emph {et~al.}(2021)\citenamefont
  {Bulancea-Lindvall}, \citenamefont {Son}, \citenamefont {Abrikosov},\ and\
  \citenamefont {Ivády}}]{bulancea-lindvall_dipolar_2021}%
  \BibitemOpen
  \bibfield  {author} {\bibinfo {author} {\bibfnamefont {O.}~\bibnamefont
  {Bulancea-Lindvall}}, \bibinfo {author} {\bibfnamefont {N.~T.}\ \bibnamefont
  {Son}}, \bibinfo {author} {\bibfnamefont {I.~A.}\ \bibnamefont {Abrikosov}},
  \ and\ \bibinfo {author} {\bibfnamefont {V.}~\bibnamefont {Ivády}},\ }\href
  {\doibase 10.1038/s41524-021-00673-8} {\bibfield  {journal} {\bibinfo
  {journal} {npj Computational Materials}\ }\textbf {\bibinfo {volume} {7}},\
  \bibinfo {pages} {1} (\bibinfo {year} {2021})}\BibitemShut {NoStop}%
\bibitem [{\citenamefont {Soltamov}\ \emph {et~al.}(2012)\citenamefont
  {Soltamov}, \citenamefont {Soltamova}, \citenamefont {Baranov},\ and\
  \citenamefont {Proskuryakov}}]{SoltamovPhysRevLett2012}%
  \BibitemOpen
  \bibfield  {author} {\bibinfo {author} {\bibfnamefont {V.~A.}\ \bibnamefont
  {Soltamov}}, \bibinfo {author} {\bibfnamefont {A.~A.}\ \bibnamefont
  {Soltamova}}, \bibinfo {author} {\bibfnamefont {P.~G.}\ \bibnamefont
  {Baranov}}, \ and\ \bibinfo {author} {\bibfnamefont {I.~I.}\ \bibnamefont
  {Proskuryakov}},\ }\href {\doibase 10.1103/PhysRevLett.108.226402} {\bibfield
   {journal} {\bibinfo  {journal} {Phys. Rev. Lett.}\ }\textbf {\bibinfo
  {volume} {108}},\ \bibinfo {pages} {226402} (\bibinfo {year}
  {2012})}\BibitemShut {NoStop}%
\bibitem [{\citenamefont {Widmann}\ \emph
  {et~al.}(2015{\natexlab{a}})\citenamefont {Widmann}, \citenamefont {Lee},
  \citenamefont {Rendler}, \citenamefont {Son}, \citenamefont {Fedder},
  \citenamefont {Paik}, \citenamefont {Yang}, \citenamefont {Zhao},
  \citenamefont {Yang}, \citenamefont {Booker}, \citenamefont {Denisenko},
  \citenamefont {Jamali}, \citenamefont {Momenzadeh}, \citenamefont {Gerhardt},
  \citenamefont {Ohshima}, \citenamefont {Gali}, \citenamefont {Janzén},\ and\
  \citenamefont {Wrachtrup}}]{widmann_coherent_2015}%
  \BibitemOpen
  \bibfield  {author} {\bibinfo {author} {\bibfnamefont {M.}~\bibnamefont
  {Widmann}}, \bibinfo {author} {\bibfnamefont {S.-Y.}\ \bibnamefont {Lee}},
  \bibinfo {author} {\bibfnamefont {T.}~\bibnamefont {Rendler}}, \bibinfo
  {author} {\bibfnamefont {N.~T.}\ \bibnamefont {Son}}, \bibinfo {author}
  {\bibfnamefont {H.}~\bibnamefont {Fedder}}, \bibinfo {author} {\bibfnamefont
  {S.}~\bibnamefont {Paik}}, \bibinfo {author} {\bibfnamefont {L.-P.}\
  \bibnamefont {Yang}}, \bibinfo {author} {\bibfnamefont {N.}~\bibnamefont
  {Zhao}}, \bibinfo {author} {\bibfnamefont {S.}~\bibnamefont {Yang}}, \bibinfo
  {author} {\bibfnamefont {I.}~\bibnamefont {Booker}}, \bibinfo {author}
  {\bibfnamefont {A.}~\bibnamefont {Denisenko}}, \bibinfo {author}
  {\bibfnamefont {M.}~\bibnamefont {Jamali}}, \bibinfo {author} {\bibfnamefont
  {S.~A.}\ \bibnamefont {Momenzadeh}}, \bibinfo {author} {\bibfnamefont
  {I.}~\bibnamefont {Gerhardt}}, \bibinfo {author} {\bibfnamefont
  {T.}~\bibnamefont {Ohshima}}, \bibinfo {author} {\bibfnamefont
  {A.}~\bibnamefont {Gali}}, \bibinfo {author} {\bibfnamefont {E.}~\bibnamefont
  {Janzén}}, \ and\ \bibinfo {author} {\bibfnamefont {J.}~\bibnamefont
  {Wrachtrup}},\ }\href {\doibase 10.1038/nmat4145} {\bibfield  {journal}
  {\bibinfo  {journal} {Nature Materials}\ }\textbf {\bibinfo {volume} {14}},\
  \bibinfo {pages} {164} (\bibinfo {year} {2015}{\natexlab{a}})}\BibitemShut
  {NoStop}%
\bibitem [{\citenamefont {Simin}\ \emph {et~al.}(2016)\citenamefont {Simin},
  \citenamefont {Soltamov}, \citenamefont {Poshakinskiy}, \citenamefont
  {Anisimov}, \citenamefont {Babunts}, \citenamefont {Tolmachev}, \citenamefont
  {Mokhov}, \citenamefont {Trupke}, \citenamefont {Tarasenko}, \citenamefont
  {Sperlich}, \citenamefont {Baranov}, \citenamefont {Dyakonov},\ and\
  \citenamefont {Astakhov}}]{Simin2016}%
  \BibitemOpen
  \bibfield  {author} {\bibinfo {author} {\bibfnamefont {D.}~\bibnamefont
  {Simin}}, \bibinfo {author} {\bibfnamefont {V.~A.}\ \bibnamefont {Soltamov}},
  \bibinfo {author} {\bibfnamefont {A.~V.}\ \bibnamefont {Poshakinskiy}},
  \bibinfo {author} {\bibfnamefont {A.~N.}\ \bibnamefont {Anisimov}}, \bibinfo
  {author} {\bibfnamefont {R.~A.}\ \bibnamefont {Babunts}}, \bibinfo {author}
  {\bibfnamefont {D.~O.}\ \bibnamefont {Tolmachev}}, \bibinfo {author}
  {\bibfnamefont {E.~N.}\ \bibnamefont {Mokhov}}, \bibinfo {author}
  {\bibfnamefont {M.}~\bibnamefont {Trupke}}, \bibinfo {author} {\bibfnamefont
  {S.~A.}\ \bibnamefont {Tarasenko}}, \bibinfo {author} {\bibfnamefont
  {A.}~\bibnamefont {Sperlich}}, \bibinfo {author} {\bibfnamefont {P.~G.}\
  \bibnamefont {Baranov}}, \bibinfo {author} {\bibfnamefont {V.}~\bibnamefont
  {Dyakonov}}, \ and\ \bibinfo {author} {\bibfnamefont {G.~V.}\ \bibnamefont
  {Astakhov}},\ }\href {\doibase 10.1103/PhysRevX.6.031014} {\bibfield
  {journal} {\bibinfo  {journal} {Phys. Rev. X}\ }\textbf {\bibinfo {volume}
  {6}},\ \bibinfo {pages} {031014} (\bibinfo {year} {2016})}\BibitemShut
  {NoStop}%
\bibitem [{\citenamefont {Riedel}\ \emph {et~al.}(2012)\citenamefont {Riedel},
  \citenamefont {Fuchs}, \citenamefont {Kraus}, \citenamefont {V\"ath},
  \citenamefont {Sperlich}, \citenamefont {Dyakonov}, \citenamefont
  {Soltamova}, \citenamefont {Baranov}, \citenamefont {Ilyin},\ and\
  \citenamefont {Astakhov}}]{Riedel2012}%
  \BibitemOpen
  \bibfield  {author} {\bibinfo {author} {\bibfnamefont {D.}~\bibnamefont
  {Riedel}}, \bibinfo {author} {\bibfnamefont {F.}~\bibnamefont {Fuchs}},
  \bibinfo {author} {\bibfnamefont {H.}~\bibnamefont {Kraus}}, \bibinfo
  {author} {\bibfnamefont {S.}~\bibnamefont {V\"ath}}, \bibinfo {author}
  {\bibfnamefont {A.}~\bibnamefont {Sperlich}}, \bibinfo {author}
  {\bibfnamefont {V.}~\bibnamefont {Dyakonov}}, \bibinfo {author}
  {\bibfnamefont {A.~A.}\ \bibnamefont {Soltamova}}, \bibinfo {author}
  {\bibfnamefont {P.~G.}\ \bibnamefont {Baranov}}, \bibinfo {author}
  {\bibfnamefont {V.~A.}\ \bibnamefont {Ilyin}}, \ and\ \bibinfo {author}
  {\bibfnamefont {G.~V.}\ \bibnamefont {Astakhov}},\ }\href {\doibase
  10.1103/PhysRevLett.109.226402} {\bibfield  {journal} {\bibinfo  {journal}
  {Phys. Rev. Lett.}\ }\textbf {\bibinfo {volume} {109}},\ \bibinfo {pages}
  {226402} (\bibinfo {year} {2012})}\BibitemShut {NoStop}%
\bibitem [{\citenamefont {Widmann}\ \emph
  {et~al.}(2015{\natexlab{b}})\citenamefont {Widmann}, \citenamefont {Lee},
  \citenamefont {Rendler}, \citenamefont {Son}, \citenamefont {Fedder},
  \citenamefont {Paik}, \citenamefont {Yang}, \citenamefont {Zhao},
  \citenamefont {Yang}, \citenamefont {Booker}, \citenamefont {Denisenko},
  \citenamefont {Jamali}, \citenamefont {Momenzadeh}, \citenamefont {Gerhardt},
  \citenamefont {Ohshima}, \citenamefont {Gali}, \citenamefont {Janz{\'e}n},\
  and\ \citenamefont {Wrachtrup}}]{Widmann2014}%
  \BibitemOpen
  \bibfield  {author} {\bibinfo {author} {\bibfnamefont {M.}~\bibnamefont
  {Widmann}}, \bibinfo {author} {\bibfnamefont {S.-Y.}\ \bibnamefont {Lee}},
  \bibinfo {author} {\bibfnamefont {T.}~\bibnamefont {Rendler}}, \bibinfo
  {author} {\bibfnamefont {N.~T.}\ \bibnamefont {Son}}, \bibinfo {author}
  {\bibfnamefont {H.}~\bibnamefont {Fedder}}, \bibinfo {author} {\bibfnamefont
  {S.}~\bibnamefont {Paik}}, \bibinfo {author} {\bibfnamefont {L.-P.}\
  \bibnamefont {Yang}}, \bibinfo {author} {\bibfnamefont {N.}~\bibnamefont
  {Zhao}}, \bibinfo {author} {\bibfnamefont {S.}~\bibnamefont {Yang}}, \bibinfo
  {author} {\bibfnamefont {I.}~\bibnamefont {Booker}}, \bibinfo {author}
  {\bibfnamefont {A.}~\bibnamefont {Denisenko}}, \bibinfo {author}
  {\bibfnamefont {M.}~\bibnamefont {Jamali}}, \bibinfo {author} {\bibfnamefont
  {S.~A.}\ \bibnamefont {Momenzadeh}}, \bibinfo {author} {\bibfnamefont
  {I.}~\bibnamefont {Gerhardt}}, \bibinfo {author} {\bibfnamefont
  {T.}~\bibnamefont {Ohshima}}, \bibinfo {author} {\bibfnamefont
  {A.}~\bibnamefont {Gali}}, \bibinfo {author} {\bibfnamefont {E.}~\bibnamefont
  {Janz{\'e}n}}, \ and\ \bibinfo {author} {\bibfnamefont {J.}~\bibnamefont
  {Wrachtrup}},\ }\href {http://dx.doi.org/10.1038/nmat4145} {\bibfield
  {journal} {\bibinfo  {journal} {Nat. Mater.}\ }\textbf {\bibinfo {volume}
  {14}},\ \bibinfo {pages} {164} (\bibinfo {year}
  {2015}{\natexlab{b}})}\BibitemShut {NoStop}%
\bibitem [{\citenamefont {Simin}\ \emph {et~al.}(2017)\citenamefont {Simin},
  \citenamefont {Kraus}, \citenamefont {Sperlich}, \citenamefont {Ohshima},
  \citenamefont {Astakhov},\ and\ \citenamefont
  {Dyakonov}}]{simin_locking_2017}%
  \BibitemOpen
  \bibfield  {author} {\bibinfo {author} {\bibfnamefont {D.}~\bibnamefont
  {Simin}}, \bibinfo {author} {\bibfnamefont {H.}~\bibnamefont {Kraus}},
  \bibinfo {author} {\bibfnamefont {A.}~\bibnamefont {Sperlich}}, \bibinfo
  {author} {\bibfnamefont {T.}~\bibnamefont {Ohshima}}, \bibinfo {author}
  {\bibfnamefont {G.~V.}\ \bibnamefont {Astakhov}}, \ and\ \bibinfo {author}
  {\bibfnamefont {V.}~\bibnamefont {Dyakonov}},\ }\href {\doibase
  10.1103/PhysRevB.95.161201} {\bibfield  {journal} {\bibinfo  {journal}
  {Physical Review B}\ }\textbf {\bibinfo {volume} {95}},\ \bibinfo {pages}
  {161201} (\bibinfo {year} {2017})}\BibitemShut {NoStop}%
\bibitem [{\citenamefont {Lee}\ \emph {et~al.}(2015)\citenamefont {Lee},
  \citenamefont {Niethammer},\ and\ \citenamefont {Wrachtrup}}]{Lee2015}%
  \BibitemOpen
  \bibfield  {author} {\bibinfo {author} {\bibfnamefont {S.-Y.}\ \bibnamefont
  {Lee}}, \bibinfo {author} {\bibfnamefont {M.}~\bibnamefont {Niethammer}}, \
  and\ \bibinfo {author} {\bibfnamefont {J.}~\bibnamefont {Wrachtrup}},\ }\href
  {\doibase 10.1103/PhysRevB.92.115201} {\bibfield  {journal} {\bibinfo
  {journal} {Phys. Rev. B}\ }\textbf {\bibinfo {volume} {92}},\ \bibinfo
  {pages} {115201} (\bibinfo {year} {2015})}\BibitemShut {NoStop}%
\bibitem [{\citenamefont {Niethammer}\ \emph {et~al.}(2016)\citenamefont
  {Niethammer}, \citenamefont {Widmann}, \citenamefont {Lee}, \citenamefont
  {Stenberg}, \citenamefont {Kordina}, \citenamefont {Ohshima}, \citenamefont
  {Son}, \citenamefont {Janz\'en},\ and\ \citenamefont
  {Wrachtrup}}]{Niethammer2016}%
  \BibitemOpen
  \bibfield  {author} {\bibinfo {author} {\bibfnamefont {M.}~\bibnamefont
  {Niethammer}}, \bibinfo {author} {\bibfnamefont {M.}~\bibnamefont {Widmann}},
  \bibinfo {author} {\bibfnamefont {S.-Y.}\ \bibnamefont {Lee}}, \bibinfo
  {author} {\bibfnamefont {P.}~\bibnamefont {Stenberg}}, \bibinfo {author}
  {\bibfnamefont {O.}~\bibnamefont {Kordina}}, \bibinfo {author} {\bibfnamefont
  {T.}~\bibnamefont {Ohshima}}, \bibinfo {author} {\bibfnamefont {N.~T.}\
  \bibnamefont {Son}}, \bibinfo {author} {\bibfnamefont {E.}~\bibnamefont
  {Janz\'en}}, \ and\ \bibinfo {author} {\bibfnamefont {J.}~\bibnamefont
  {Wrachtrup}},\ }\href {\doibase 10.1103/PhysRevApplied.6.034001} {\bibfield
  {journal} {\bibinfo  {journal} {Phys. Rev. Applied}\ }\textbf {\bibinfo
  {volume} {6}},\ \bibinfo {pages} {034001} (\bibinfo {year}
  {2016})}\BibitemShut {NoStop}%
\bibitem [{\citenamefont {Anisimov}\ \emph {et~al.}(2016)\citenamefont
  {Anisimov}, \citenamefont {Simin}, \citenamefont {Soltamov}, \citenamefont
  {Lebedev}, \citenamefont {Baranov}, \citenamefont {Astakhov},\ and\
  \citenamefont {Dyakonov}}]{Anisimov2016}%
  \BibitemOpen
  \bibfield  {author} {\bibinfo {author} {\bibfnamefont {A.~N.}\ \bibnamefont
  {Anisimov}}, \bibinfo {author} {\bibfnamefont {D.}~\bibnamefont {Simin}},
  \bibinfo {author} {\bibfnamefont {V.~A.}\ \bibnamefont {Soltamov}}, \bibinfo
  {author} {\bibfnamefont {S.~P.}\ \bibnamefont {Lebedev}}, \bibinfo {author}
  {\bibfnamefont {P.~G.}\ \bibnamefont {Baranov}}, \bibinfo {author}
  {\bibfnamefont {G.~V.}\ \bibnamefont {Astakhov}}, \ and\ \bibinfo {author}
  {\bibfnamefont {V.}~\bibnamefont {Dyakonov}},\ }\href
  {http://dx.doi.org/10.1038/srep33301} {\bibfield  {journal} {\bibinfo
  {journal} {Sci. Rep.}\ }\textbf {\bibinfo {volume} {6}},\ \bibinfo {pages}
  {33301} (\bibinfo {year} {2016})}\BibitemShut {NoStop}%
\bibitem [{\citenamefont {Kraus}\ \emph {et~al.}(2014)\citenamefont {Kraus},
  \citenamefont {Soltamov}, \citenamefont {Riedel}, \citenamefont {V\"ath},
  \citenamefont {Fuchs}, \citenamefont {Sperlich}, \citenamefont {Baranov},
  \citenamefont {Dyakonov},\ and\ \citenamefont {Astakhov}}]{Kraus2014}%
  \BibitemOpen
  \bibfield  {author} {\bibinfo {author} {\bibfnamefont {H.}~\bibnamefont
  {Kraus}}, \bibinfo {author} {\bibfnamefont {V.~A.}\ \bibnamefont {Soltamov}},
  \bibinfo {author} {\bibfnamefont {D.}~\bibnamefont {Riedel}}, \bibinfo
  {author} {\bibfnamefont {S.}~\bibnamefont {V\"ath}}, \bibinfo {author}
  {\bibfnamefont {F.}~\bibnamefont {Fuchs}}, \bibinfo {author} {\bibfnamefont
  {A.}~\bibnamefont {Sperlich}}, \bibinfo {author} {\bibfnamefont {P.~G.}\
  \bibnamefont {Baranov}}, \bibinfo {author} {\bibfnamefont {V.}~\bibnamefont
  {Dyakonov}}, \ and\ \bibinfo {author} {\bibfnamefont {G.~V.}\ \bibnamefont
  {Astakhov}},\ }\href {\doibase 10.1038/nphys2826} {\bibfield  {journal}
  {\bibinfo  {journal} {Nat. Phys.}\ }\textbf {\bibinfo {volume} {10}},\
  \bibinfo {pages} {157} (\bibinfo {year} {2014})}\BibitemShut {NoStop}%
\bibitem [{\citenamefont {Nagy}\ \emph {et~al.}(2019)\citenamefont {Nagy},
  \citenamefont {Niethammer}, \citenamefont {Widmann}, \citenamefont {Chen},
  \citenamefont {Udvarhelyi}, \citenamefont {Bonato}, \citenamefont {Hassan},
  \citenamefont {Karhu}, \citenamefont {Ivanov}, \citenamefont {Son},
  \citenamefont {Maze}, \citenamefont {Ohshima}, \citenamefont {Soykal},
  \citenamefont {Gali}, \citenamefont {Lee}, \citenamefont {Kaiser},\ and\
  \citenamefont {Wrachtrup}}]{nagy_high-fidelity_2019}%
  \BibitemOpen
  \bibfield  {author} {\bibinfo {author} {\bibfnamefont {R.}~\bibnamefont
  {Nagy}}, \bibinfo {author} {\bibfnamefont {M.}~\bibnamefont {Niethammer}},
  \bibinfo {author} {\bibfnamefont {M.}~\bibnamefont {Widmann}}, \bibinfo
  {author} {\bibfnamefont {Y.-C.}\ \bibnamefont {Chen}}, \bibinfo {author}
  {\bibfnamefont {P.}~\bibnamefont {Udvarhelyi}}, \bibinfo {author}
  {\bibfnamefont {C.}~\bibnamefont {Bonato}}, \bibinfo {author} {\bibfnamefont
  {J.~U.}\ \bibnamefont {Hassan}}, \bibinfo {author} {\bibfnamefont
  {R.}~\bibnamefont {Karhu}}, \bibinfo {author} {\bibfnamefont {I.~G.}\
  \bibnamefont {Ivanov}}, \bibinfo {author} {\bibfnamefont {N.~T.}\
  \bibnamefont {Son}}, \bibinfo {author} {\bibfnamefont {J.~R.}\ \bibnamefont
  {Maze}}, \bibinfo {author} {\bibfnamefont {T.}~\bibnamefont {Ohshima}},
  \bibinfo {author} {\bibfnamefont {O.~O.}\ \bibnamefont {Soykal}}, \bibinfo
  {author} {\bibfnamefont {A.}~\bibnamefont {Gali}}, \bibinfo {author}
  {\bibfnamefont {S.-Y.}\ \bibnamefont {Lee}}, \bibinfo {author} {\bibfnamefont
  {F.}~\bibnamefont {Kaiser}}, \ and\ \bibinfo {author} {\bibfnamefont
  {J.}~\bibnamefont {Wrachtrup}},\ }\href {\doibase 10.1038/s41467-019-09873-9}
  {\bibfield  {journal} {\bibinfo  {journal} {Nature Communications}\ }\textbf
  {\bibinfo {volume} {10}},\ \bibinfo {pages} {1954} (\bibinfo {year}
  {2019})}\BibitemShut {NoStop}%
\bibitem [{\citenamefont {Son}\ \emph {et~al.}(2020)\citenamefont {Son},
  \citenamefont {Anderson}, \citenamefont {Bourassa}, \citenamefont {Miao},
  \citenamefont {Babin}, \citenamefont {Widmann}, \citenamefont {Niethammer},
  \citenamefont {Ul~Hassan}, \citenamefont {Morioka}, \citenamefont {Ivanov},
  \citenamefont {Kaiser}, \citenamefont {Wrachtrup},\ and\ \citenamefont
  {Awschalom}}]{son_developing_2020}%
  \BibitemOpen
  \bibfield  {author} {\bibinfo {author} {\bibfnamefont {N.~T.}\ \bibnamefont
  {Son}}, \bibinfo {author} {\bibfnamefont {C.~P.}\ \bibnamefont {Anderson}},
  \bibinfo {author} {\bibfnamefont {A.}~\bibnamefont {Bourassa}}, \bibinfo
  {author} {\bibfnamefont {K.~C.}\ \bibnamefont {Miao}}, \bibinfo {author}
  {\bibfnamefont {C.}~\bibnamefont {Babin}}, \bibinfo {author} {\bibfnamefont
  {M.}~\bibnamefont {Widmann}}, \bibinfo {author} {\bibfnamefont
  {M.}~\bibnamefont {Niethammer}}, \bibinfo {author} {\bibfnamefont
  {J.}~\bibnamefont {Ul~Hassan}}, \bibinfo {author} {\bibfnamefont
  {N.}~\bibnamefont {Morioka}}, \bibinfo {author} {\bibfnamefont {I.~G.}\
  \bibnamefont {Ivanov}}, \bibinfo {author} {\bibfnamefont {F.}~\bibnamefont
  {Kaiser}}, \bibinfo {author} {\bibfnamefont {J.}~\bibnamefont {Wrachtrup}}, \
  and\ \bibinfo {author} {\bibfnamefont {D.~D.}\ \bibnamefont {Awschalom}},\
  }\href {\doibase 10.1063/5.0004454} {\bibfield  {journal} {\bibinfo
  {journal} {Applied Physics Letters}\ }\textbf {\bibinfo {volume} {116}},\
  \bibinfo {pages} {190501} (\bibinfo {year} {2020})}\BibitemShut {NoStop}%
\bibitem [{\citenamefont {Babin}\ \emph {et~al.}(2022)\citenamefont {Babin},
  \citenamefont {Stöhr}, \citenamefont {Morioka}, \citenamefont {Linkewitz},
  \citenamefont {Steidl}, \citenamefont {Wörnle}, \citenamefont {Liu},
  \citenamefont {Hesselmeier}, \citenamefont {Vorobyov}, \citenamefont
  {Denisenko}, \citenamefont {Hentschel}, \citenamefont {Gobert}, \citenamefont
  {Berwian}, \citenamefont {Astakhov}, \citenamefont {Knolle}, \citenamefont
  {Majety}, \citenamefont {Saha}, \citenamefont {Radulaski}, \citenamefont
  {Son}, \citenamefont {Ul-Hassan}, \citenamefont {Kaiser},\ and\ \citenamefont
  {Wrachtrup}}]{babin_fabrication_2022}%
  \BibitemOpen
  \bibfield  {author} {\bibinfo {author} {\bibfnamefont {C.}~\bibnamefont
  {Babin}}, \bibinfo {author} {\bibfnamefont {R.}~\bibnamefont {Stöhr}},
  \bibinfo {author} {\bibfnamefont {N.}~\bibnamefont {Morioka}}, \bibinfo
  {author} {\bibfnamefont {T.}~\bibnamefont {Linkewitz}}, \bibinfo {author}
  {\bibfnamefont {T.}~\bibnamefont {Steidl}}, \bibinfo {author} {\bibfnamefont
  {R.}~\bibnamefont {Wörnle}}, \bibinfo {author} {\bibfnamefont
  {D.}~\bibnamefont {Liu}}, \bibinfo {author} {\bibfnamefont {E.}~\bibnamefont
  {Hesselmeier}}, \bibinfo {author} {\bibfnamefont {V.}~\bibnamefont
  {Vorobyov}}, \bibinfo {author} {\bibfnamefont {A.}~\bibnamefont {Denisenko}},
  \bibinfo {author} {\bibfnamefont {M.}~\bibnamefont {Hentschel}}, \bibinfo
  {author} {\bibfnamefont {C.}~\bibnamefont {Gobert}}, \bibinfo {author}
  {\bibfnamefont {P.}~\bibnamefont {Berwian}}, \bibinfo {author} {\bibfnamefont
  {G.~V.}\ \bibnamefont {Astakhov}}, \bibinfo {author} {\bibfnamefont
  {W.}~\bibnamefont {Knolle}}, \bibinfo {author} {\bibfnamefont
  {S.}~\bibnamefont {Majety}}, \bibinfo {author} {\bibfnamefont
  {P.}~\bibnamefont {Saha}}, \bibinfo {author} {\bibfnamefont {M.}~\bibnamefont
  {Radulaski}}, \bibinfo {author} {\bibfnamefont {N.~T.}\ \bibnamefont {Son}},
  \bibinfo {author} {\bibfnamefont {J.}~\bibnamefont {Ul-Hassan}}, \bibinfo
  {author} {\bibfnamefont {F.}~\bibnamefont {Kaiser}}, \ and\ \bibinfo {author}
  {\bibfnamefont {J.}~\bibnamefont {Wrachtrup}},\ }\href {\doibase
  10.1038/s41563-021-01148-3} {\bibfield  {journal} {\bibinfo  {journal}
  {Nature Materials}\ }\textbf {\bibinfo {volume} {21}},\ \bibinfo {pages} {67}
  (\bibinfo {year} {2022})}\BibitemShut {NoStop}%
\bibitem [{\citenamefont {Soltamov}\ \emph {et~al.}(2019)\citenamefont
  {Soltamov}, \citenamefont {Kasper}, \citenamefont {Poshakinskiy},
  \citenamefont {Anisimov}, \citenamefont {Mokhov}, \citenamefont {Sperlich},
  \citenamefont {Tarasenko}, \citenamefont {Baranov}, \citenamefont
  {Astakhov},\ and\ \citenamefont {Dyakonov}}]{soltamov_excitation_2019}%
  \BibitemOpen
  \bibfield  {author} {\bibinfo {author} {\bibfnamefont {V.~A.}\ \bibnamefont
  {Soltamov}}, \bibinfo {author} {\bibfnamefont {C.}~\bibnamefont {Kasper}},
  \bibinfo {author} {\bibfnamefont {A.~V.}\ \bibnamefont {Poshakinskiy}},
  \bibinfo {author} {\bibfnamefont {A.~N.}\ \bibnamefont {Anisimov}}, \bibinfo
  {author} {\bibfnamefont {E.~N.}\ \bibnamefont {Mokhov}}, \bibinfo {author}
  {\bibfnamefont {A.}~\bibnamefont {Sperlich}}, \bibinfo {author}
  {\bibfnamefont {S.~A.}\ \bibnamefont {Tarasenko}}, \bibinfo {author}
  {\bibfnamefont {P.~G.}\ \bibnamefont {Baranov}}, \bibinfo {author}
  {\bibfnamefont {G.~V.}\ \bibnamefont {Astakhov}}, \ and\ \bibinfo {author}
  {\bibfnamefont {V.}~\bibnamefont {Dyakonov}},\ }\href {\doibase
  10.1038/s41467-019-09429-x} {\bibfield  {journal} {\bibinfo  {journal}
  {Nature Communications}\ }\textbf {\bibinfo {volume} {10}},\ \bibinfo {pages}
  {1678} (\bibinfo {year} {2019})}\BibitemShut {NoStop}%
\bibitem [{\citenamefont {Ramsay}\ and\ \citenamefont
  {Rossi}(2020)}]{ramsay_relaxation_2020}%
  \BibitemOpen
  \bibfield  {author} {\bibinfo {author} {\bibfnamefont {A.~J.}\ \bibnamefont
  {Ramsay}}\ and\ \bibinfo {author} {\bibfnamefont {A.}~\bibnamefont {Rossi}},\
  }\href {\doibase 10.1103/PhysRevB.101.165307} {\bibfield  {journal} {\bibinfo
   {journal} {Physical Review B}\ }\textbf {\bibinfo {volume} {101}},\ \bibinfo
  {pages} {165307} (\bibinfo {year} {2020})}\BibitemShut {NoStop}%
\bibitem [{\citenamefont {Gugler}\ \emph {et~al.}(2018)\citenamefont {Gugler},
  \citenamefont {Astner}, \citenamefont {Angerer}, \citenamefont
  {Schmiedmayer}, \citenamefont {Majer},\ and\ \citenamefont
  {Mohn}}]{gugler_ab_2018}%
  \BibitemOpen
  \bibfield  {author} {\bibinfo {author} {\bibfnamefont {J.}~\bibnamefont
  {Gugler}}, \bibinfo {author} {\bibfnamefont {T.}~\bibnamefont {Astner}},
  \bibinfo {author} {\bibfnamefont {A.}~\bibnamefont {Angerer}}, \bibinfo
  {author} {\bibfnamefont {J.}~\bibnamefont {Schmiedmayer}}, \bibinfo {author}
  {\bibfnamefont {J.}~\bibnamefont {Majer}}, \ and\ \bibinfo {author}
  {\bibfnamefont {P.}~\bibnamefont {Mohn}},\ }\href {\doibase
  10.1103/PhysRevB.98.214442} {\bibfield  {journal} {\bibinfo  {journal}
  {Physical Review B}\ }\textbf {\bibinfo {volume} {98}},\ \bibinfo {pages}
  {214442} (\bibinfo {year} {2018})}\BibitemShut {NoStop}%
\bibitem [{\citenamefont {Park}\ \emph {et~al.}(2020)\citenamefont {Park},
  \citenamefont {Zhou},\ and\ \citenamefont
  {Bernardi}}]{park_spin-phonon_2020}%
  \BibitemOpen
  \bibfield  {author} {\bibinfo {author} {\bibfnamefont {J.}~\bibnamefont
  {Park}}, \bibinfo {author} {\bibfnamefont {J.-J.}\ \bibnamefont {Zhou}}, \
  and\ \bibinfo {author} {\bibfnamefont {M.}~\bibnamefont {Bernardi}},\ }\href
  {\doibase 10.1103/PhysRevB.101.045202} {\bibfield  {journal} {\bibinfo
  {journal} {Physical Review B}\ }\textbf {\bibinfo {volume} {101}},\ \bibinfo
  {pages} {045202} (\bibinfo {year} {2020})}\BibitemShut {NoStop}%
\bibitem [{\citenamefont {Xu}\ \emph {et~al.}(2020)\citenamefont {Xu},
  \citenamefont {Habib}, \citenamefont {Kumar}, \citenamefont {Wu},
  \citenamefont {Sundararaman},\ and\ \citenamefont
  {Ping}}]{xu_spin-phonon_2020}%
  \BibitemOpen
  \bibfield  {author} {\bibinfo {author} {\bibfnamefont {J.}~\bibnamefont
  {Xu}}, \bibinfo {author} {\bibfnamefont {A.}~\bibnamefont {Habib}}, \bibinfo
  {author} {\bibfnamefont {S.}~\bibnamefont {Kumar}}, \bibinfo {author}
  {\bibfnamefont {F.}~\bibnamefont {Wu}}, \bibinfo {author} {\bibfnamefont
  {R.}~\bibnamefont {Sundararaman}}, \ and\ \bibinfo {author} {\bibfnamefont
  {Y.}~\bibnamefont {Ping}},\ }\href {\doibase 10.1038/s41467-020-16063-5}
  {\bibfield  {journal} {\bibinfo  {journal} {Nature Communications}\ }\textbf
  {\bibinfo {volume} {11}},\ \bibinfo {pages} {2780} (\bibinfo {year}
  {2020})}\BibitemShut {NoStop}%
\bibitem [{\citenamefont {Iv\'ady}(2020)}]{IvadyPRb2020}%
  \BibitemOpen
  \bibfield  {author} {\bibinfo {author} {\bibfnamefont {V.}~\bibnamefont
  {Iv\'ady}},\ }\href {\doibase 10.1103/PhysRevB.101.155203} {\bibfield
  {journal} {\bibinfo  {journal} {Phys. Rev. B}\ }\textbf {\bibinfo {volume}
  {101}},\ \bibinfo {pages} {155203} (\bibinfo {year} {2020})}\BibitemShut
  {NoStop}%
\bibitem [{\citenamefont {Ivády}\ \emph {et~al.}(2021)\citenamefont {Ivády},
  \citenamefont {Zheng}, \citenamefont {Wickenbrock}, \citenamefont {Bougas},
  \citenamefont {Chatzidrosos}, \citenamefont {Nakamura}, \citenamefont
  {Sumiya}, \citenamefont {Ohshima}, \citenamefont {Isoya}, \citenamefont
  {Budker}, \citenamefont {Abrikosov},\ and\ \citenamefont
  {Gali}}]{ivady_photoluminescence_2021}%
  \BibitemOpen
  \bibfield  {author} {\bibinfo {author} {\bibfnamefont {V.}~\bibnamefont
  {Ivády}}, \bibinfo {author} {\bibfnamefont {H.}~\bibnamefont {Zheng}},
  \bibinfo {author} {\bibfnamefont {A.}~\bibnamefont {Wickenbrock}}, \bibinfo
  {author} {\bibfnamefont {L.}~\bibnamefont {Bougas}}, \bibinfo {author}
  {\bibfnamefont {G.}~\bibnamefont {Chatzidrosos}}, \bibinfo {author}
  {\bibfnamefont {K.}~\bibnamefont {Nakamura}}, \bibinfo {author}
  {\bibfnamefont {H.}~\bibnamefont {Sumiya}}, \bibinfo {author} {\bibfnamefont
  {T.}~\bibnamefont {Ohshima}}, \bibinfo {author} {\bibfnamefont
  {J.}~\bibnamefont {Isoya}}, \bibinfo {author} {\bibfnamefont
  {D.}~\bibnamefont {Budker}}, \bibinfo {author} {\bibfnamefont {I.~A.}\
  \bibnamefont {Abrikosov}}, \ and\ \bibinfo {author} {\bibfnamefont
  {A.}~\bibnamefont {Gali}},\ }\href {\doibase 10.1103/PhysRevB.103.035307}
  {\bibfield  {journal} {\bibinfo  {journal} {Physical Review B}\ }\textbf
  {\bibinfo {volume} {103}},\ \bibinfo {pages} {035307} (\bibinfo {year}
  {2021})}\BibitemShut {NoStop}%
\bibitem [{\citenamefont {Iv\'ady}\ \emph {et~al.}(2017)\citenamefont
  {Iv\'ady}, \citenamefont {Davidsson}, \citenamefont {Son}, \citenamefont
  {Ohshima}, \citenamefont {Abrikosov},\ and\ \citenamefont
  {Gali}}]{IvadyVSi-4H}%
  \BibitemOpen
  \bibfield  {author} {\bibinfo {author} {\bibfnamefont {V.}~\bibnamefont
  {Iv\'ady}}, \bibinfo {author} {\bibfnamefont {J.}~\bibnamefont {Davidsson}},
  \bibinfo {author} {\bibfnamefont {N.~T.}\ \bibnamefont {Son}}, \bibinfo
  {author} {\bibfnamefont {T.}~\bibnamefont {Ohshima}}, \bibinfo {author}
  {\bibfnamefont {I.~A.}\ \bibnamefont {Abrikosov}}, \ and\ \bibinfo {author}
  {\bibfnamefont {A.}~\bibnamefont {Gali}},\ }\href {\doibase
  10.1103/PhysRevB.96.161114} {\bibfield  {journal} {\bibinfo  {journal} {Phys.
  Rev. B}\ }\textbf {\bibinfo {volume} {96}},\ \bibinfo {pages} {161114}
  (\bibinfo {year} {2017})}\BibitemShut {NoStop}%
\bibitem [{\citenamefont {Iv{\'a}dy}\ \emph {et~al.}(2018)\citenamefont
  {Iv{\'a}dy}, \citenamefont {Abrikosov},\ and\ \citenamefont
  {Gali}}]{IvadyNPJCM2018}%
  \BibitemOpen
  \bibfield  {author} {\bibinfo {author} {\bibfnamefont {V.}~\bibnamefont
  {Iv{\'a}dy}}, \bibinfo {author} {\bibfnamefont {I.~A.}\ \bibnamefont
  {Abrikosov}}, \ and\ \bibinfo {author} {\bibfnamefont {A.}~\bibnamefont
  {Gali}},\ }\href {\doibase 10.1038/s41524-018-0132-5} {\bibfield  {journal}
  {\bibinfo  {journal} {npj Computational Materials}\ }\textbf {\bibinfo
  {volume} {4}},\ \bibinfo {pages} {76} (\bibinfo {year} {2018})}\BibitemShut
  {NoStop}%
\bibitem [{\citenamefont {Falk}\ \emph {et~al.}(2013)\citenamefont {Falk},
  \citenamefont {Buckley}, \citenamefont {Calusine}, \citenamefont {Koehl},
  \citenamefont {Dobrovitski}, \citenamefont {Politi}, \citenamefont {Zorman},
  \citenamefont {Feng},\ and\ \citenamefont {Awschalom}}]{falk_polytype_2013}%
  \BibitemOpen
  \bibfield  {author} {\bibinfo {author} {\bibfnamefont {A.~L.}\ \bibnamefont
  {Falk}}, \bibinfo {author} {\bibfnamefont {B.~B.}\ \bibnamefont {Buckley}},
  \bibinfo {author} {\bibfnamefont {G.}~\bibnamefont {Calusine}}, \bibinfo
  {author} {\bibfnamefont {W.~F.}\ \bibnamefont {Koehl}}, \bibinfo {author}
  {\bibfnamefont {V.~V.}\ \bibnamefont {Dobrovitski}}, \bibinfo {author}
  {\bibfnamefont {A.}~\bibnamefont {Politi}}, \bibinfo {author} {\bibfnamefont
  {C.~A.}\ \bibnamefont {Zorman}}, \bibinfo {author} {\bibfnamefont {P.~X.-L.}\
  \bibnamefont {Feng}}, \ and\ \bibinfo {author} {\bibfnamefont {D.~D.}\
  \bibnamefont {Awschalom}},\ }\href {\doibase 10.1038/ncomms2854} {\bibfield
  {journal} {\bibinfo  {journal} {Nature Communications}\ }\textbf {\bibinfo
  {volume} {4}},\ \bibinfo {pages} {1819} (\bibinfo {year} {2013})}\BibitemShut
  {NoStop}%
\bibitem [{\citenamefont {Bulancea-Lindvall}\ \emph {et~al.}(2022)\citenamefont
  {Bulancea-Lindvall}, \citenamefont {Eiles}, \citenamefont {Son},
  \citenamefont {Abrikosov},\ and\ \citenamefont {Iv\'{a}dy}}]{oscar_VSi_2}%
  \BibitemOpen
  \bibfield  {author} {\bibinfo {author} {\bibfnamefont {O.}~\bibnamefont
  {Bulancea-Lindvall}}, \bibinfo {author} {\bibfnamefont {M.~T.}\ \bibnamefont
  {Eiles}}, \bibinfo {author} {\bibfnamefont {N.~T.}\ \bibnamefont {Son}},
  \bibinfo {author} {\bibfnamefont {I.~A.}\ \bibnamefont {Abrikosov}}, \ and\
  \bibinfo {author} {\bibfnamefont {V.}~\bibnamefont {Iv\'{a}dy}},\ }\href@noop
  {} {\  (\bibinfo {year} {2022})},\ \bibinfo {note} {{\bf To be
  submitted}}\BibitemShut {NoStop}%
\bibitem [{\citenamefont {Singh}\ \emph {et~al.}(2020)\citenamefont {Singh},
  \citenamefont {Anisimov}, \citenamefont {Nagalyuk}, \citenamefont {Mokhov},
  \citenamefont {Baranov},\ and\ \citenamefont
  {Suter}}]{singh_experimental_2020}%
  \BibitemOpen
  \bibfield  {author} {\bibinfo {author} {\bibfnamefont {H.}~\bibnamefont
  {Singh}}, \bibinfo {author} {\bibfnamefont {A.~N.}\ \bibnamefont {Anisimov}},
  \bibinfo {author} {\bibfnamefont {S.~S.}\ \bibnamefont {Nagalyuk}}, \bibinfo
  {author} {\bibfnamefont {E.~N.}\ \bibnamefont {Mokhov}}, \bibinfo {author}
  {\bibfnamefont {P.~G.}\ \bibnamefont {Baranov}}, \ and\ \bibinfo {author}
  {\bibfnamefont {D.}~\bibnamefont {Suter}},\ }\href {\doibase
  10.1103/PhysRevB.101.134110} {\bibfield  {journal} {\bibinfo  {journal}
  {Physical Review B}\ }\textbf {\bibinfo {volume} {101}},\ \bibinfo {pages}
  {134110} (\bibinfo {year} {2020})}\BibitemShut {NoStop}%
\end{thebibliography}

%

\end{document}